\documentclass{pas}

\usepackage{multirow}
\usepackage{natbib}
\usepackage{url}
\usepackage{booktabs}  

\usepackage{graphicx} 
\usepackage{stmaryrd}
\usepackage{graphicx}
\usepackage[font={small}]{subcaption}
\usepackage[font=small,labelfont=bf]{caption}

\renewenvironment{thebibliography}[1]{%
 \thebib@list
}{
 \endlist
}

\def\thebib@list{%
 \list{\null}{%
 \partopsep 0mm
 \leftmargin 1.2em
 \labelsep 0mm
 \itemindent -1.2em
 \itemsep 0.0\baselineskip
 \parsep 0mm
  \usecounter{enumi}%
 }%
 
}%
\usepackage[letterspace=100]{microtype}

\begin{document}

\lefttitle{Publications of the Astronomical Society of Australia}
\righttitle{Trevor Butrum et al.}

\jnlPage{1}{10}
\jnlDoiYr{2025}
\doival{10.1017/pasa.xxxx.xx}

\title{\textbf{A comparison of dust content and properties in GAMA/G10-COSMOS/3D-HST and \textsc{Simba} cosmological simulations}}

\author{\sn{Trevor} \gn{Butrum},$^{1\star}$ \sn{Benne} \gn{Holwerda},$^1$ \sn{Romeel} \gn{Davé},$^2$, \sn{Kyle} \gn{Cook},$^1$ \sn{Clayton} \gn{Robertson},$^1$ and \sn{Jochen} \gn{Liske}$^3$}

\affil{$^1$ Department of Physics and Astronomy, University of Louisville, 102 Natural Sciences Building, Louisville, KY 40292, USA,\newline$^2$ Institute for Astronomy, Royal Observatory, Univ. of Edinburgh, Edinburgh EH9 3HJ, UK, \newline$^3$ Hamburger Sternwarte, Universität Hamburg, Gojenbergsweg 112, 21029 Hamburg, Germany}

\corresp{Trevor Butrum, Email: tbutrum@trevor-astronomy.com}

\citeauth{Trevor B. and Benne H. (2024), A comparison of dust content in GAMA/G10-COSMOS/3D-HST and SIMBA cosmological simulations. {\it Publications of the Astronomical Society of Australia} {\bf 00}, 1--10. https://doi.org/10.1017/pasa.xxxx.xx}

\history{(Received xx xx xxxx; revised xx xx xxxx; accepted xx xx xxxx)}

\begin{abstract}
The abundance of dust within galaxies directly influences their evolution. Contemporary models attempt to match this abundance by simulating the processes of dust creation, growth, and destruction. While these models are accurate, they require refinement, especially at earlier epochs. This study aims to compare simulated and observed datasets and identify discrepancies between the two, providing a basis for future improvements. We utilise simulation data from the \textsc{Simba} cosmological simulation suite and observed data from the Galaxy and Mass Assembly (GAMA), a subset of the Cosmic Evolution Survey (G10-COSMOS), and the Hubble Space Telescope (3D-HST). We selected galaxies in the observed and simulated data in a stellar mass range of (\(10^{8.59} < M_\odot < 10^{11.5}\)) and at redshift bins centering around \(z = 0.0\), \(z = 0.1\), \(z = 0.5\), \(z = 1.0\), and \(z = 1.5\) in a homogeneous dust mass range (\(10^{6} < M_D [M_\odot] < 10^{9}\)). Our results show notable deviations between \textsc{Simba} and observed data for dust-poor and rich galaxies, with strong indications that differences in galaxy populations and \textsc{Simba} limitations are the underlying cause rather than the dust physics implemented in \textsc{Simba} itself.


\end{abstract}


\maketitle

\section{Introduction}

In modern galaxy evolution studies, dust is crucial in how galaxies evolve and change across cosmic time. The abundance of dust in galaxies is directly connected with galaxy evolution (\citealp{Santini14}), as stars would not form effectively without it, as dust catalyses the formation of molecules \citep{Wakelam17, Chen18} and enables the fragmentation of gas clouds \citep{Omukai05, Schneider06}. However, the exact abundance of this dust in galaxies is unclear, especially at earlier epochs when the age of the Universe was comparable to the typical dust formation time scales \citep{Todini01, Bianchi07, Leśniewska19}.

Modern galaxy evolution models have attempted to match the different dust abundances across galaxy populations at different epochs \citep{McKinnon16, McKinnon17a, Popping17, Hou19, Triani20, Vijayan19, Parente22a, Parente23}. Although significant advancements have been made in modelling the life cycle of dust, the mechanisms behind the creation, growth, and destruction of dust, particularly concerning dust content, remain a topic of debate in the modern era \citep[][]{Hensley23, Ragone-Figueroa24, Yates24}. Enhancing our models through direct comparisons of simulated and observed data is crucial for deepening our understanding of dust's life cycle and its contents in galaxies and, consequently, galaxy evolution. Recent results suggest a substantial evolution of the dust content of galaxies since $z\sim1$ \citep{Parente23, Eales24}.


Despite these current efforts, most comparisons currently lack a sufficiently large sample size to reveal significant discrepancies, particularly in earlier epochs where surveys are lacking. Most modern comparisons with models and observations \citep[e.g.]{Li19d, McKinnon17a, Popping17} rely on low sample size observational datasets nearly a decade old \citep{Eales09, Dunne11, Clemens13} that focus on later epochs (\(0 < z < 1\)) containing 82, 1867, and 234 sources, respectively. The only outlier most studies utilise is \cite{Beeston18}, which includes 15,750 sources at a redshift \(< 0.1\) due to its utilisation of more modern surveys. However, this is only at a small redshift range at later epochs. To properly understand the properties that govern the abundance of dust in galaxies, more complete samples at earlier epochs are necessary for enhancing our comparisons.

The current aim of this study is to compare the simulated and observed datasets with a larger sample size and later redshifts than previously accomplished to help remedy this issue. We strive to identify the discrepancies between the dust mass generated by modern galaxy evolution models and measured observational data. Using an extensive data set to identify these discrepancies, this paper serves as a first-effort basis for improving galaxy evolution models across epochs up to \(z=1.5\).

\begin{figure*}[htb] 
\centering
\begin{subfigure}{0.5\textwidth}
\graphicspath{ {InitialPlots/}}
\centering
\includegraphics[width=1.0\textwidth]{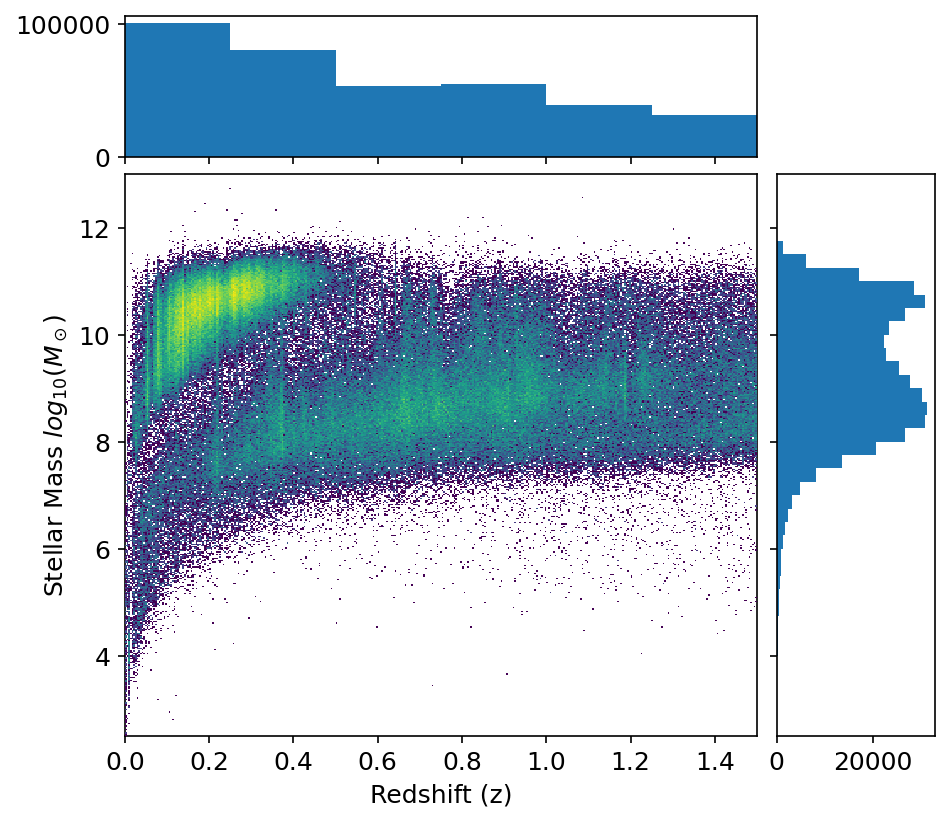}
\label{StellarRaw}
\end{subfigure}%
\begin{subfigure}{0.5\textwidth}
\graphicspath{ {InitialPlots/}}
\centering
\includegraphics[width=1.0\textwidth]{"Joined_Catalog_Dust_Mass_Raw_Data"}
\label{DustRaw}
\end{subfigure}
\caption{The complete data sets of GAMA, G10-COSMOS, 3D-HST with stellar mass (left) and dust mass (right) up till the intermediate Universe (\(z\approx 1.5\)). The more yellow the data displayed on the figures, the greater the data contained in that area, and the bluer the less. The clumps observed correspond to different surveys. GAMA covers up to the near Universe (\(z < 0.5\)) and corresponds to the high-density clump at the top left of the plots. G10-COSMOS and 3D-HST cover up to the intermediate Universe (\(z < 1.75\)) and are harder to distinguish as there is little separation between the data.}

\label{Fig1}
\end{figure*}


We, therefore, utilise the dust mass data produced by the hydrodynamical model \textsc{Simba} and compare it using various methods with a large, homogeneous, and detailed dust mass dataset, as described in \cite{Driver18}. The observed dataset includes data from Galaxy and Mass Assembly (GAMA), an offshoot of GAMA, the Cosmic Evolution Survey (G10-COSMOS), and the Hubble Space Telescope 3D project (3D-HST). We note that the data received from G10-COSMOS and \textsc{Simba} were obtained via private communication and are not publicly available.

The paper is organised as follows: in Section \ref{Section2}, we summarise our observational datasets: GAMA, G10-COSMOS, and 3D-HST. In Section \ref{SIMBADATA}, we briefly describe the dust evolution model present in \textsc{Simba} and the simulations used in our paper. In Section \ref{Section4}, we explain our galaxy selection progress for both the observational and simulated datasets. In Section \ref{Results}, we explore the results of our comparisons between the two datasets, and in Section \ref{discussion}, we discuss further the implications of these results and put them into context with other works. Finally, we finish with our conclusions in Section \ref{conclusion}.

\section{Data} \label{Section2}
We combine the three datasets outlined in \cite{Driver18}: GAMA \citep{Driver11, Liske15}, G10-COSMOS \citep{Davies15b, Andrews17}, and 3D-HST \citep{Momcheva16a}. These studies contain data from the ultraviolet (UV), mid-infrared (MIR), and far-IR (FIR) wavelengths. The GAMA and G10-COSMOS studies also contain constraints from \textit{Herschel Space Observatory's} SPIRE \citep{Griffin10, Poglitsch10} and PACS instruments \citep{Eales10c, Oliver12}, which allows for robust dust mass measurements. The three studies GAMA, G10-COSMOS, and 3D-HST, extend to the nearby (\(z \leq 0.5\)), intermediate (\(z < 1.75\)), and high-\(z\) Universe (\(z < 5.0\)), respectively. Each dataset comprises approximately 200,000 galaxies that sample a broad range in stellar mass, dust mass, and look-back time.  Each dataset was subject to data cuts based on flux limitations set in \cite{Driver18} and active galactic nuclei (AGN) contamination removal. Each sample was processed using \textsc{magphys}, a spectral energy distribution fitting (SED) code, to provide estimates of dust mass, stellar mass, and star-formation rates (SFR) based on the flux limitations  (see Sections \ref{GAMA}, \ref{G10-COSMOS}, and \ref{3D-HST}).


\subsection{GAMA} \label{GAMA}
The GAMA survey \citep{Driver09, Driver11, Liske15} is a complete (\(98\) \% to \(r < 19.8\) mag) spectroscopic survey up to the near universe (\(z \leq 0.5\)). The survey consists of five regions G02 (\(\sim 56\) deg\(^2\)), G09 (\(\sim 60\) deg\(^2\)), G12 (\(\sim 60\) deg\(^2\)), G15 (\(\sim 60\) deg\(^2\)), and G23 (\(\sim 51\) deg\(^2\)) for a total area of roughly 180 deg\(^2\) with about 70,000 galaxies per region. \citealp[]{Driver18} uses LAMBDARCatv01, a catalogue for GAMA made with in-house software (\textsc{lambdar};  \citealp{Wright16}) which provides flux limits with their errors, upper-limits, and flags. These objects were filtered from an initial value of 200,246 to 128,568 based on specific quality criteria, with values subsequently adjusted to be compatible with \textsc{magphys}. They also implemented an additional cut step to eliminate high stellar mass outliers. We use the \textsc{magphys} SED fits data-products with the high stellar mass cut (\citealp[]{Driver18}) for the described 128,568 objects in this study.

\begin{figure*}[!htb] 
\centering
\begin{subfigure}{0.6\textwidth}
\graphicspath{ {FinalPlots/}}
\centering
\includegraphics[width=\textwidth]{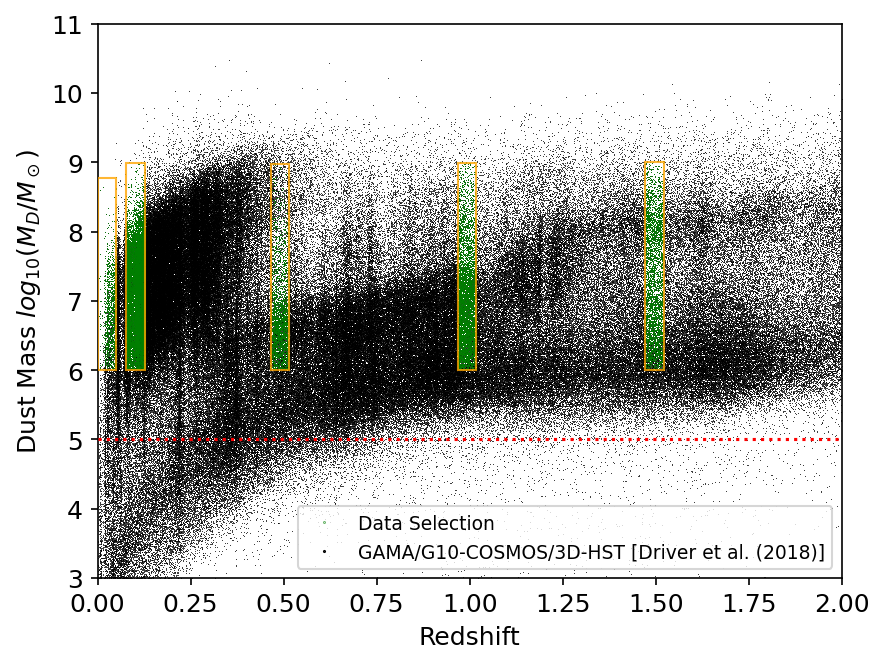}
\label{f:fullM:MD:plot}
\centering

\end{subfigure}%

\caption{The dust mass ranges selected for this study. We plot GAMA/G10-COSMOS/3D-HST data as black dots. We plot the selected data with the ranges applied in green dots with a box surrounding them. The red dotted line represents the dust mass volume limits of the surveys. Note that we exclude GAMA data at \(z=0.5\) due to volume-related issues in the selections.}
\label{Fig2}
\end{figure*}

\subsection{G10-COSMOS} \label{G10-COSMOS}
The G10-COSMOS survey \citep{Davies15b, Andrews17} is a subset of \textit{HST} COSMOS survey that covers a 1 deg\(^2\) region of the survey \citep{Scoville07a, Scoville07b} and is 100 percent redshift complete up to a specified flux limit  of \(i < 25\) mag. The survey includes wavelengths from UV to FIR. The survey uses the data collected by the Galaxy Evolution Explorer (\textit{GALEX}; \citealp{Martin05}; UV), Canada–France–Hawaii Telescope (CFHT; \citealp{Cuillandre12}; optical), Subaru (\citealp{Taniguchi07}; optical), \textit{HST} (\citealp{Scoville07a}; optical), Visible and Infrared Survey Telescope (VISTA; \citealp{Emerson02, Jarvis13}; NIR), \textit{Spitzer} (\citealp{Werner04, Sanders07}; MIR), and some \textit{Herschel} (\citealp{Pilbratt03, Shirley21}; FIR). 

Similar to the GAMA dataset, G10-COSMOS utilizes the \textsc{lambdar} G10CosmosLAMBDARCatv06 catalog \citep{Andrews17a,Wright16}. This catalogue also similarly underwent AGN cuts based on different criteria outlined \citep{Donely12, Seymour08, Laigle16}, as well as the removal of high stellar mass outliers. It includes a total of 142,260 objects found in the intermediate Universe (\(z < 1.75\)). We use the \textsc{magphys} SED fits data products with the AGN cut (\citealp[]{Driver18}) for the 142,260 objects described in this study. 

\subsection{3D-HST} \label{3D-HST}
The 3D-HST survey (\citealp{Brammer12, Momcheva16a}) is an almost complete (up to 85 percent at \(F160W = 26.0\) mag) photometric, grating-prism (GPRISM), and spectroscopic survey up to the high-z universe (\(z < 5\)) that covers a \(\sim0.2\) deg\(^2\) area of the sky. The fields 3D-HST covers are sub-regions of the All-Wavelength Extended Groth Strip International Survey (AEGIS) \citep{Davis07}, Cosmic Evolution Survey (COSMOS) \citep{Scoville07a}, Great Observatories Origins Deep Survey (GOODS-South) \citep{Nonino09}, GOODS-North \citep{Barger08}, UKIRT Infrared Deep Sky Survey (UKIDSSUDS) \cite{Almaini07}, and Cosmic Assembly Near-infrared Deep Extragalactic Legacy Survey (CANDELS) \citep{Grogin11}.  Unlike GAMA and G10-COSMOS, 3D-HST does not include FIR coverage from \textit{Herschel}, which is currently under development \citep{Hurley17}. The 3D-HST catalog consists of 204,294 galaxies and AGN, with a redshift estimate of 3,839 from spectroscopic, 15,518 from GRISM, and 185,843 from photometric data. We use the \textsc{magphys} SED fits data products from the fair AGN cut, which follows the criteria set by \cite{Donely12}, as well as the exclusion of high stellar mass outliers (\citealp[]{Driver18}), for the 188,235 objects discussed in this study.

\begin{table*}[!htb]
\label{Table1}
\begin{center}

\caption{Results of our sample selections. Data with prevalent volume limitations have been removed and labelled as "\dots". }

\begin{tabular}{c c c c c c} 
\hline
 {\hspace{1.75em}} Redshift Bins ($z$) {\hspace{1.75em}}   & {\hspace{1.75em}} GAMA {\hspace{1.75em}} &  {\hspace{1.75em}} G10-COSMOS {\hspace{1.75em}} & {\hspace{1.75em}} 3D-HST{\hspace{1.75em}} & {\hspace{1.75em}} Total {\hspace{1.75em}} & {\hspace{1.75em}}\textsc{Simba} {\hspace{1.75em}}
 \\ 
  \hline

  0.0-0.05  & 1411  & \dots & \dots & 1411  & 28 984  \\ 

  0.07748-0.12748  & 14 949   & 175   & \dots & 15 124 & 27 715  \\

  0.46544-0.51544  & \dots   & 2276     & 406 & 2682  & 21 247 \\

  0.9677-1.0177  & \dots  & 4616   & 1543 & 6176  & 15 713\\

  1.4717-1.5217  & \dots  & 2034   & 1661 & 3695 &  11 808   \\
 \hline
Total & 16 360 & 9101 & 3610 & 29 088  & 105 467
 \\
 \hline
 \label{Table1}
\end{tabular}

\end{center}
\end{table*}

\subsection{\textsc{magphys}
} \label{magphys}

To calculate the physical properties for dust mass, stellar mass, and SFR used in this study, the three datasets containing 128,568, 142,260, and 188,235 objects, respectively, were parsed through the SED-fitting code \textsc{magphys}. For further details on the data preparation process and \textsc{magphys}, we point the reader to \citealp{Driver18} and \citealp{Da-Cunha08}. This parsing through \textsc{magphys} standardises the assumptions and systematics of dust mass, stellar mass, and SFR estimates between the datasets, which minimises the errors between the surveys and allows us to combine the datasets and make proper comparisons with the simulated galaxies in \textsc{Simba}.

\section{The \textsc{Simba} simulation} \label{SIMBADATA}

In this study, we utilise the cosmological simulation \textsc{Simba} \citep{Dave19}, which is the successor to \textsc{mufasa} and is built upon the \textsc{gizmo} hydrodynamic code (\citealp{Hopkins15}), for our comparisons. Here, we will present a summary of the relevant physics in \textsc{Simba} to help explain our findings in Section \ref{Results}.

\textsc{Simba} presents a self-consistent model for dust production, growth, and destruction \citep{Li19d, Dave19}. This model tracks eleven elements (H, He, C, N, O, Ne, Mg, Si, S, Ca, and Fe) across cosmic time, with enrichment sourced from Type II SNe, Type Ia SNe, and AGB stars. In \textsc{Simba}, dust is fully coupled with gas flows, a treatment considered accurate due to the simulation's under-resolved drag force and radiative pressure. Dust grains are also modeled with a consistent size of \(0.1 \mu \text{m}\) and a density of \(2.4 \, \text{g cm}^{-3}\) \citep{Draine03}. 

Dust production in \textsc{Simba} is determined by taking fixed fractions of metals from Type II SNe and AGB star condensations, based on the work of \cite{Popping17}, which updates earlier findings by \cite{Dwek98}. Below are the equations (outlined in \citealp{Li19d, Dave19}) that describe how dust mass is determined using AGB stars and Type II SNe where \(m_{i,d}^{\text{j}}\) refers to the \textit{i}th element (C, O, Mg, Si, S, Ca, and Fe) produced by the \textit{j}th stellar process (AGB stars or SNeII) and where \(m_{i,ej}^{\text{j}}\) refers to the mass ejected from the \textit{j}th process.

The mass of dust produced by AGB stars with a carbon-to-oxygen ratio of greater than one (\(C/O > 1\)) is expressed as 

\begin{equation}
    m_{i,d}^{\text{AGB}} =
    \begin{cases}
        \delta_C^{\text{AGB}} \left( m_{C,ej}^{\text{AGB}} - 0.75 m_{O,ej}^{\text{AGB}} \right), & i = C \\
        0, & \text{otherwise},
    \end{cases}
\end{equation}

where \(\delta_i^{\text{AGB}}\) is the fixed condensation efficiency of element \textit{i} in AGB stars based on \citealp{Ferrarotti06}. The dust mass produced by AGB stars with carbon-to-oxygen ratios less than one (\(C/O < 1\)) is expressed as 

\begin{equation}
    m_{i,d}^{\text{AGB}} =
    \begin{cases}
        0, & i = C \\
        16 \sum\limits_{i=\text{Mg,Si,S,Ca,Fe}} \delta_i^{\text{AGB}} m_{i,ej}^{\text{AGB}}, & i = O \\
        \delta_i^{\text{AGB}} m_{i,ej}^{\text{AGB}}, & \text{otherwise},
    \end{cases}
\end{equation}

where \(\mu_i\) is the mass of element i. Finally, the mass of dust produced by Type II SNe is described as 

\begin{equation}
    m_{i,d}^{\text{SNII}} =
    \begin{cases}
        \delta_C^{\text{SNII}} m_{C,ej}^{\text{SNII}}, & i = C \\
        16 \sum\limits_{i=\text{Mg,Si,S,Ca,Fe}} \delta_i^{\text{SNII}} m_{i,ej}^{\text{SNII}}, & i = O \\
        \delta_i^{\text{SNII}} m_{i,ej}^{\text{SNII}}, & \text{otherwise},
    \end{cases}
\end{equation}

where \(\delta_i^{\text{SNII}}\) is the fixed condensation efficiency of element i for SNe II based on \citealp{Bianchi07}.

Dust grains are then formed through the accretion of local gaseous metals following \cite{Dwek98}.

\begin{equation}
    \left(\frac{dM_d}{dt}\right)_{grow} = \left(1 - \frac{M_d}{M_{metal}}\right)\left(\frac{M_d}{\tau_{accr}}\right)
\end{equation}

where \(M_{metal}\) is the total mass of dust and local gas-phase metals. The accretion timescale \(\tau_{accr}\) is then found following \cite{Hirashita00} and \cite{Asano13} and is

\begin{equation}
    \tau_{accr} = \tau_{ref}\left(\frac{\rho_{ref}}{\rho_g}\right)\left(\frac{T_{ref}}{T_g}\right) \left(\frac{Z_{M_\odot}}{Z_g}\right)
\end{equation}

where \(p_g\), \(T_g\), and \(Z_g\) are the local gas density, temperature, and metallicity, respectively, and the others are reference values. \(p_{ref}=100\)H atoms cm\(^{-3}\), \(T_{ref} = 20\)K and \(\tau_{ref} = 10\) Myr.

These dust grains are eventually destroyed by thermal sputtering following \cite{Popping17} and \cite{ McKinnon17a}, with the timescale expressed as

\begin{align}
    \tau_{sp} &= a\left|\frac{da}{dt}\right|^{-1} \sim (0.17\text{Gyr})\left(\frac{a}{0.1\mu\text{m}}\right)\left(\frac{10^{-27}\text{g cm}^{-3}}{\rho_g}\right) \notag \\
    &\times \left[\left(\frac{T_0}{T_g}\right)^{w} + 1\right].
\end{align}

where \(w=2.5\) controls the low-temperature scaling of the sputtering and \(T_0=2\times10^6\text{K}\) is the temperature above where the sputtering curve starts to flatten. The growth rate of the dust due to this sputtering is then calculated by

\begin{equation}
    \left(\frac{dM_d}{dt}\right)_{sp}=-\frac{M_d}{\tau_{sp}/3}.
\end{equation}

SN blasts are not resolved in the simulations and are implemented by a subgrid model for dust destruction by SN shocks following \cite{Dwek80}, \cite{Seab83}, and \cite{McKee87, McKee89}. The timescale \(\tau_{de}\) is

\begin{equation}
    \tau_{de} = \frac{M_g}{\epsilon\gamma M_s},
\end{equation}

where \(M_g\) is the local gas mass, \(\epsilon=0.3\) is the efficiency of local gain destruction, \(\gamma\) is the local SNe II rate, and \(M_g\) is the mass of the local gas shocked at \(100\text{km s}^{-1}\). Finally, a solar abundance of \(Z_{\odot} = 0.0134\) is assumed for the star formation and grain growth models taken from \cite{Asplund09}. 

The parameters governing these processes are listed in Table 1 \citep{Li19d}. It is important to note that these parameters are adjusted to align with observations at low redshifts; therefore, the \(z=0\) simulation matches observations by construction rather than prediction.

We utilize the \textsc{Simba} m100n1024 simulations at redshifts \(z = 0.0, 0.1, 0.5, 1.0, 1.5\) for our comparisons. This simulated cosmological cube has a side length of \(100h^{-1}\) Mpc and contains \(1024^3\) dark matter and gas elements. The mass resolution limits of \textsc{Simba} are \(9.6 \times 10^7 M_\odot\) for dark matter particles and \(1.82 \times 10^7 M_\odot\) for gas elements. This simulation adheres to the Planck16 concordant cosmology method \citep{Planck-Collaboration16}.

\section{Sample selection}\label{Section4}

To start our comparisons between the observations and \textsc{Simba}, we must ensure that our combined observational dataset is comprehensive and addresses any limitations inherent to each dataset. To accomplish this, we applied specific restrictions regarding redshift, dust mass, stellar mass, and SFR to our datasets. These measures helped facilitate fair and accurate comparisons between observations and simulations.  

Our implementation of redshift restrictions relies primarily on the snapshots provided by \textsc{Simba}, as described in Section \ref{SIMBADATA}. Accordingly, we also must limit our observational datasets to these ranges. A cut of \(z=0.025\) was then applied at each redshift on either side of the initial redshift to ensure consistent comparisons across cosmic time, resulting in a total bin width of \(\Delta z = 0.05\). Since no data is available before \(z = 0.0\), we have limited the right-hand side of the data to the specified redshift bin while maintaining the same range. 

These datasets also have limitations concerning volume and sensitivity (see Section \ref{limitations}). Given these constraints, along with the stellar mass and dust mass resolution limits of \textsc{Simba} described in Section \ref{SIMBADATA}, we chose to limit our dust mass to \(10^{6} < M_D [M_\odot] < 10^{9}\), stellar mass to \(10^{8.59} < M_\odot < 10^{11.5}\), and SFR to \(10^{-2} < M_{\odot/\text{yr}} < 10^{2}\). These selected ranges proved to be the most effective in minimising the limitations of each dataset and ensuring a complete sample across the datasets. We also excluded GAMA at \(z=0.5\), G10-COSMOS at \(z=0.0\), and 3D-HST for \(z<0.5\) because these volume limitations at these redshifts negatively influenced our results.

Please refer to Table \ref{Table1} and Figure \ref{Fig2} for a comprehensive overview of our redshift selections and a quantitative comparison of the number of galaxies between observations and simulations.

\section{Results} \label{Results}

After selecting and constraining our samples, we compare the dust content and properties between GAMA/G10-COSMOS/3D-HST and \textsc{Simba}, following the selections outlined in Section \ref{Section4}.  We acknowledge that the significant differences in sample size (Table \ref{Table1}) may introduce some bias in our results and could account for some observed differences between the observations and the model.

\begin{figure}[htbp] 
\centering
\begin{subfigure}{0.43\textwidth}
\graphicspath{ {FinalPlots/}}
\centering
a\includegraphics[width=1\textwidth]{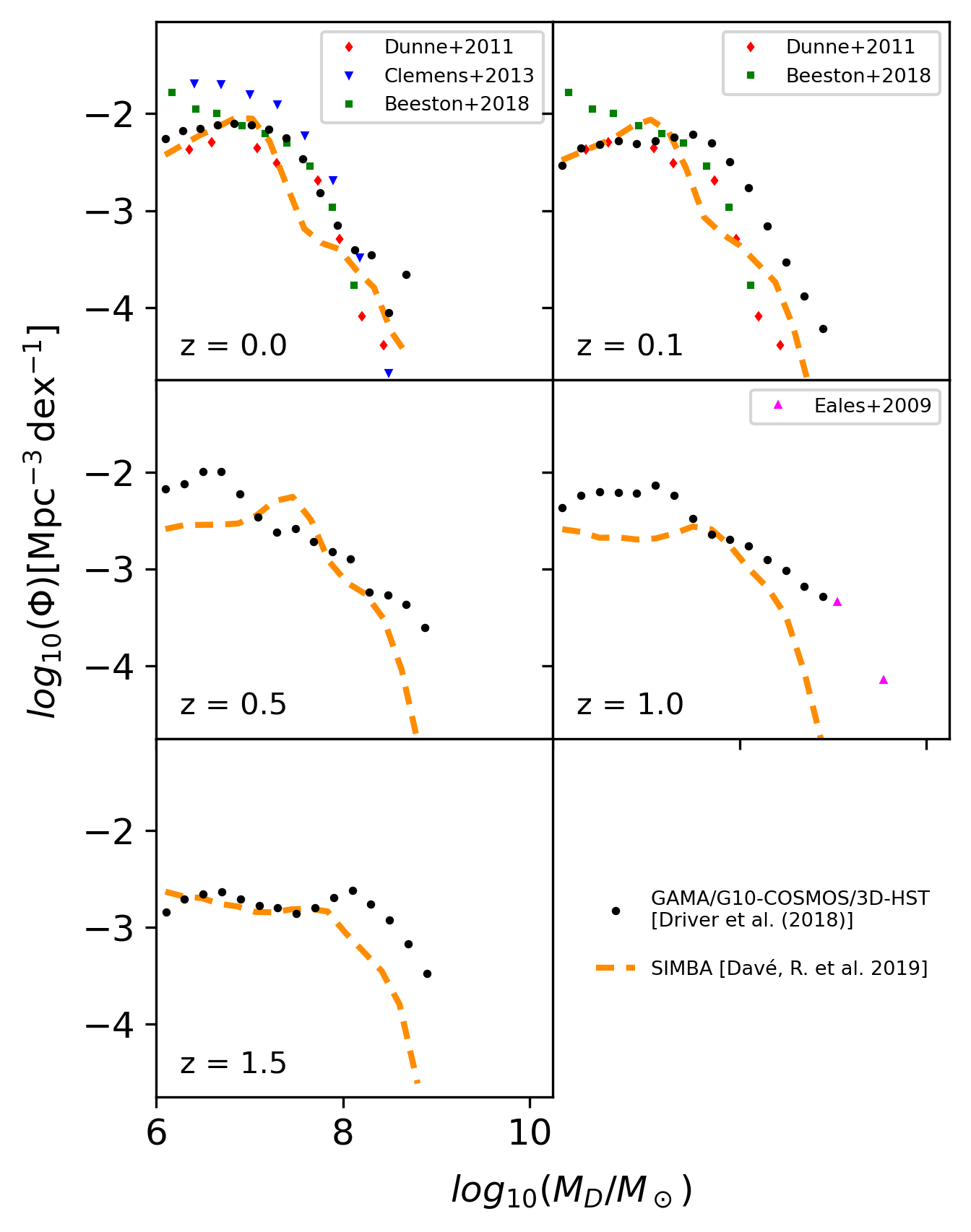}
\label{dmfs}
\end{subfigure}%
\caption{DMFs from observations and simulations at \(z = 0-1.5\). \cite{Eales09} is plotted from data in the range \(0.6 < z < 1.0\) and \cite{Dunne11} and \cite{Beeston18} is plotted from data in the range \(0.0 < z < 0.1\). Our results are not standardised to the cosmological parameters described in \cite{Li19d}.} 
\label{Fig5}
\end{figure}

\begin{figure}[htbp] 
\centering
\begin{subfigure}{0.48\textwidth}
\graphicspath{ {FinalPlots/}}
\centering
\includegraphics[width=1\textwidth]{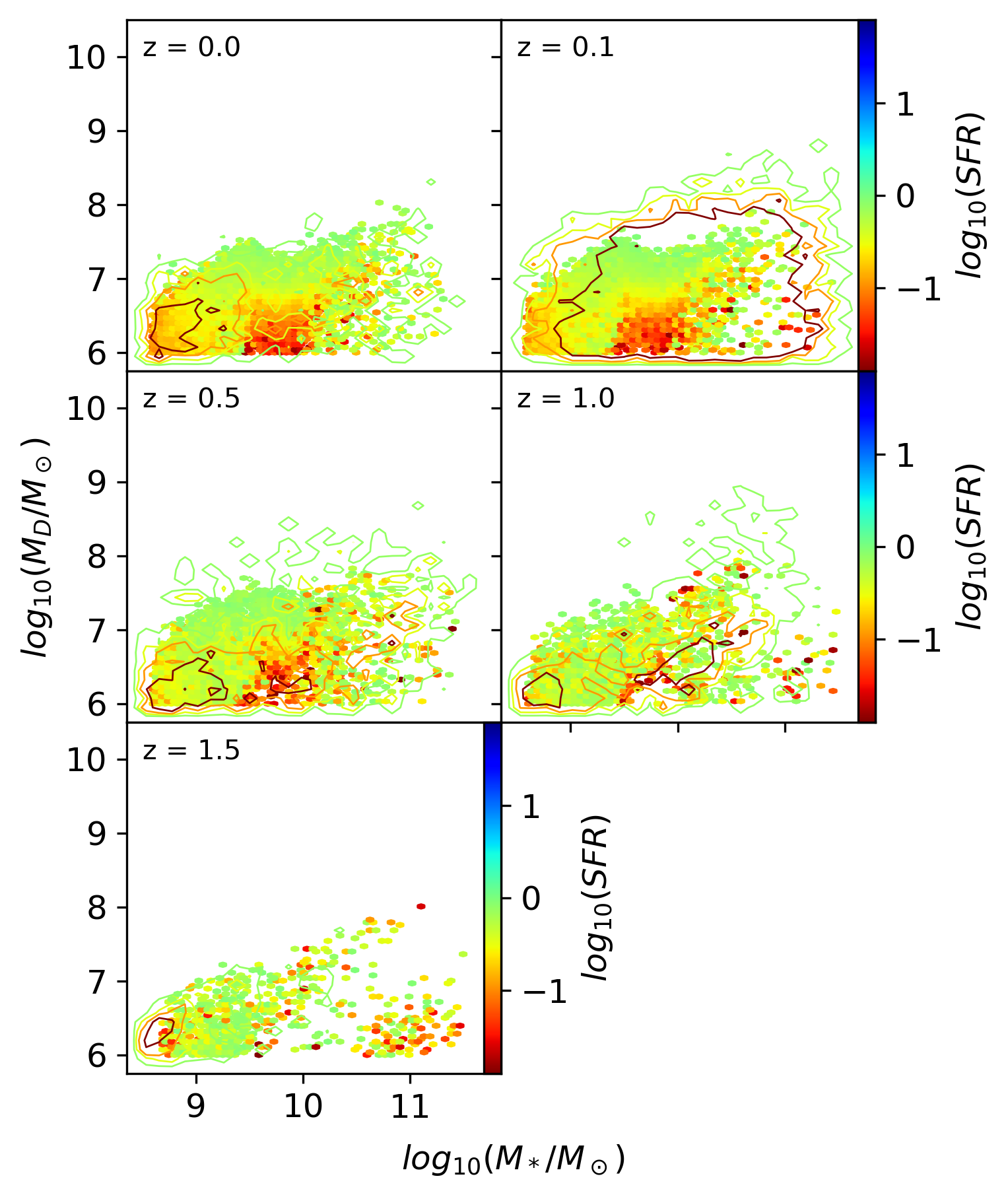}
\end{subfigure}%
\caption{A relation between dust mass and stellar mass of \textsc{Simba} and GAMA/G10-COSMOS/3D-HST of quenching galaxies (SFR \(\leq\) 0)  at \(z=0.0-1.5\).   Observations, represented by contours weighted by star formation rates, are drawn at -8, -4, -2, -0.5, 0.5, 2, 4, and 8. The simulation, \textsc{Simba}, is plotted as hex-bins. Both are colour-coded according to star formation rates.}
\label{obs_dsm}
\end{figure}

\subsection{Dust mass functions}
\label{DMF}

\begin{figure}[htpb]
\centering
\begin{subfigure}{0.48\textwidth}
\graphicspath{ {FinalPlots/}}
\centering
\includegraphics[width=1\textwidth]{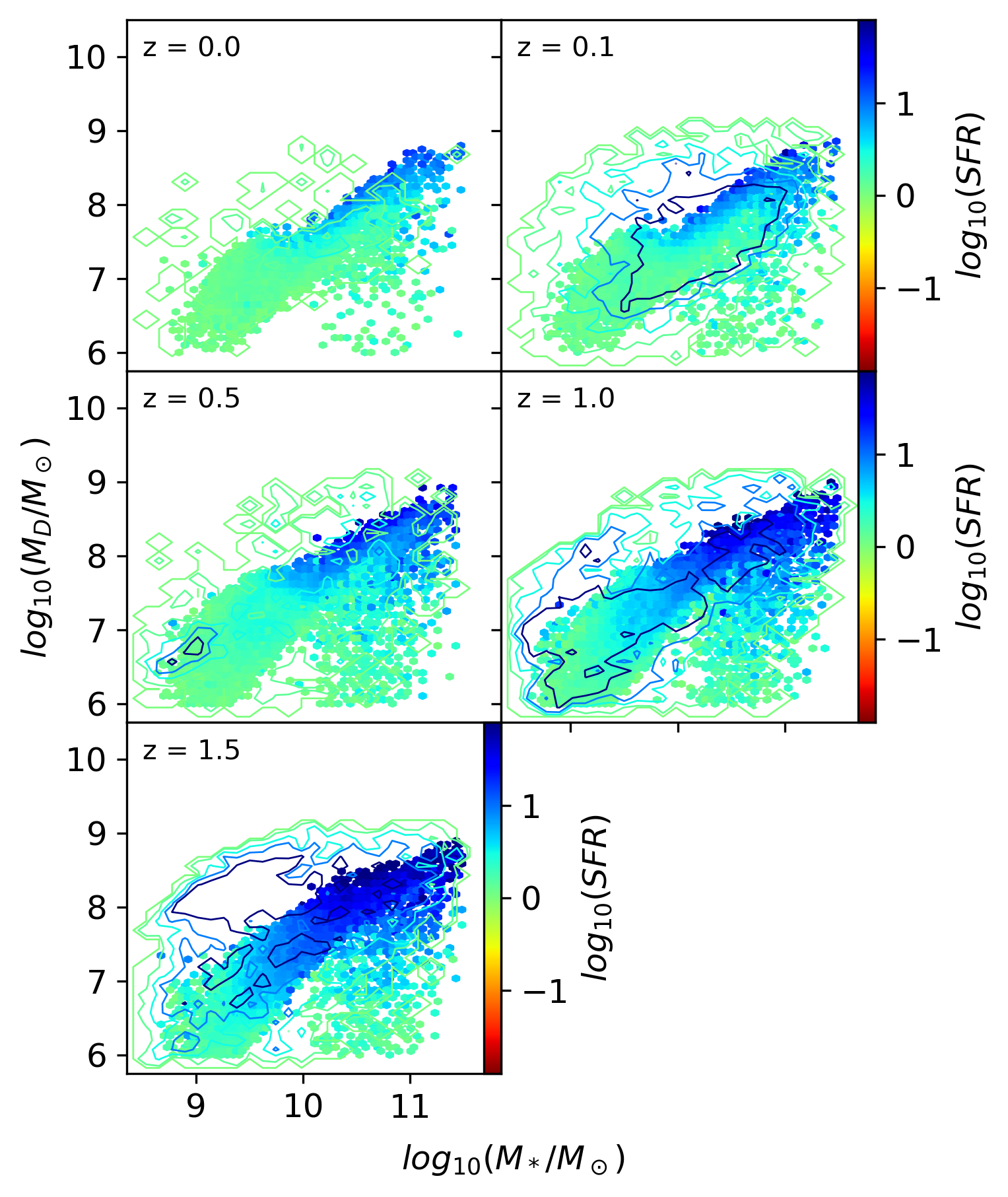}
\end{subfigure}%
\caption{A relation between dust mass and stellar mass of \textsc{Simba} and GAMA/G10-COSMOS/3D-HST of star-forming galaxies (SFR \(\geq\) 0)   at \(z=0.0-1.5\).  Observations, represented by contours weighted by star formation rates, are drawn at -8, -4, -2, -0.5, 0.5, 2, 4, and 8. The simulation, \textsc{Simba}, is plotted as hex-bins. Both are colour-coded according to star formation rates.}
\label{sim_dsm}
\end{figure}

Figure \ref{Fig5} shows the redshift evolution of the DMFs, comparing ours, the simulation and other observational data consisting of data from \cite{Dunne11}, \cite{Clemens13}, and \cite{Beeston18} at \(z=0.0\) and \cite{Eales09} at \(z=1.0\). We use these findings, presented in \cite{Li19d}, as an additional complement to our results. We also do not standardise our data to their assumed dust mass absorption coefficient and cosmological parameters, as the difference between the observations and simulations is trivial. This may introduce some slight discrepancies, but it should be sufficient for our comparisons.

At \(z=0.0\), our data agrees well with \textsc{Simba} and broadly follows the other observed data. \textsc{Simba} data underestimates the DMF in the medium-mass range, where the observational dataset contains about \(\sim36\%\) of the total galaxies observed. Otherwise, it matches quite well in the low-mass end. The model at \(z=0.1\) exhibits similar characteristics with a more significant underestimate in the medium-mass range where a substantial portion of the observational dataset lies (\(\sim50\%\)) and a slight underestimation near the high-mass end, but this range consists only of \(\sim5\%\) of the total galaxies in the observations. The model at \(z=0.5\) is where Simba begins to deviate from the observational data noticeably. \textsc{Simba} significantly underestimates the low-mass end where a significant portion of the observational data lies (\(\sim74\%\)) and slightly underestimates the high-mass end, but similar to \(z=0.1\), it consists of only about \(\sim5\%\) of the total observational total. This pattern continues in the \(z=1.0\) model, where \textsc{Simba} underestimates dust near the low and high ends again, where the observational data constitutes \(\sim52\%\) and \(\sim9\%\) of the total galaxies, respectively. In contrast, the \(z=1.5\) model largely agrees near the low-mass end and similarly underestimates the high end. However, unlike \(z=0.1\) and \(z=0.5\), the high end of the observational dataset contains about \(\sim25\%\) of the total galaxies. Please refer to Table \ref{tab:galaxy_counts} for the complete counts of the observational data in each redshift, dust mass, and stellar mass bin. We will explore the stellar and dust mass bins further in the next section.

\subsection{Dust mass versus stellar mass}

\begin{figure}[htbp] 
\centering
\begin{subfigure}{0.48\textwidth}
\graphicspath{ {FinalPlots/}}
\centering
\includegraphics[width=1\textwidth]{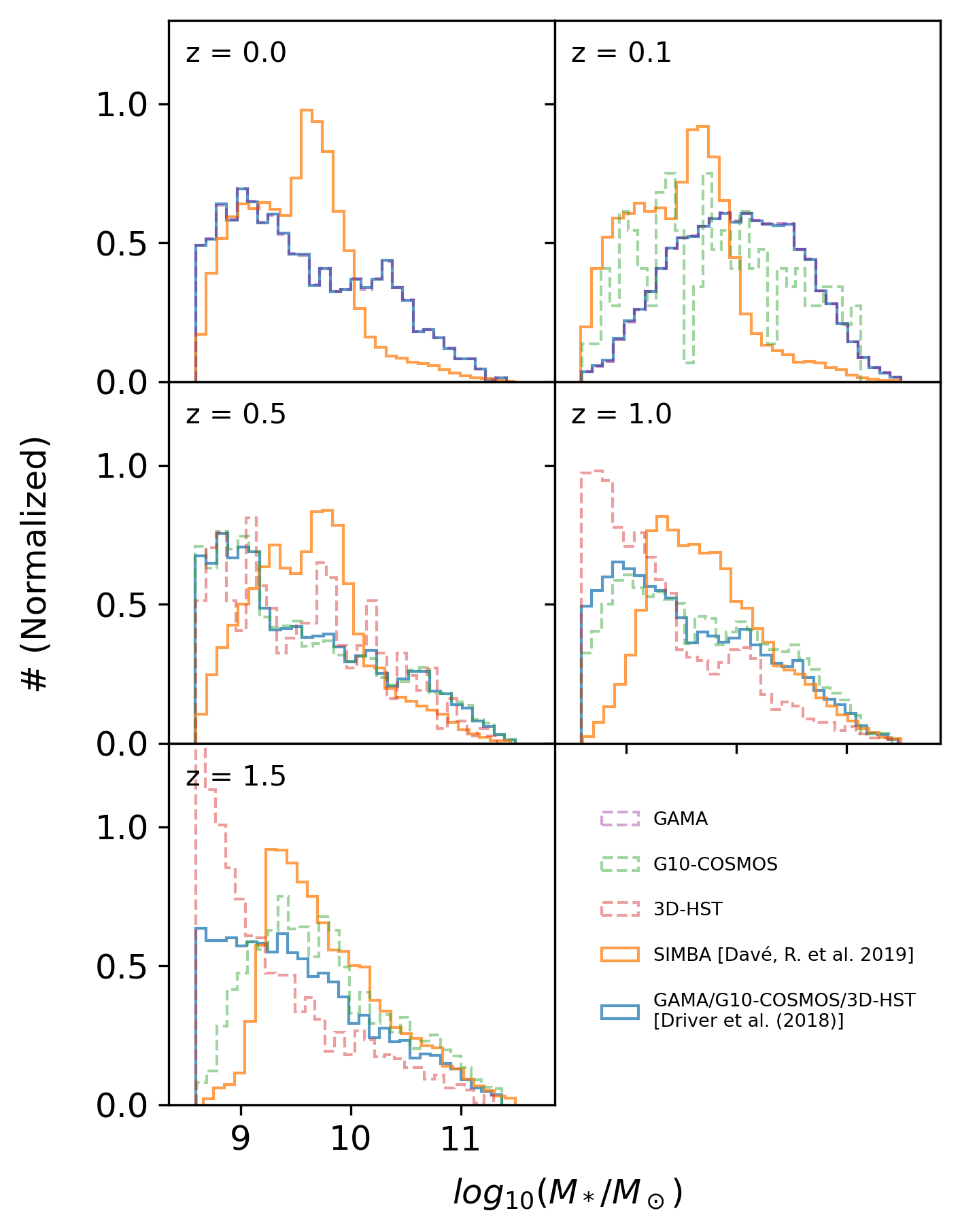}
\end{subfigure}%
\caption{Normalized counts of stellar mass from \textsc{Simba} and GAMA/G10-COSMOS/3D-HST. The observational dataset is separated into individual surveys to highlight the distinctions between them.}
\label{stellar_mass_norm}
\end{figure}

To further investigate the inconsistencies shown in Figure \ref{Fig5}, we will examine the evolution of dust with stellar mass throughout cosmic time between the two datasets. We illustrate our comparisons in Figure \ref{obs_dsm} (log(SFR) \(\leq\) 0)  and \ref{sim_dsm} (log(SFR) \(\geq\) 0)   as hex-bin (simulations)  and contour (observations)  plots color-coded with the star formation rates of the datasets and normalised stellar mass in Figure \ref{stellar_mass_norm} to examine the populations explicitly. The graphs (Figure \ref{obs_dsm} and \ref{sim_dsm}) are divided into log(SFR) \(\leq\) 0 and log(SFR) \(\geq\) 0 galaxies and catalogues (Figure \ref{stellar_mass_norm})  to highlight where the differences between the datasets are the most concentrated. 

We will start with our results from Figure \ref{obs_dsm} representing low-SFR galaxies. At \(z=0\), our data generally agrees well with \textsc{Simba}. However, there is a noticeable mismatch between the observations and the simulations. The observations' low-SFR dust-poor galaxies are found at the lowest stellar masses (\(10^{8.59}-10^{9.5}\)) consisting of about \(74\%\) of the total observed galaxies in this SFR range, while in the simulations, they are more prevalent at medium stellar masses (\(10^{9.5}-10^{10.5}\)). However, compared to \textsc{Simba}, the sample size of GAMA (see Table \ref{Table1}) is significantly smaller, which may explain this difference.  In contrast to \(z=0.0\), at \(z=0.1\), the data matches closely with the model but misses massive dust-poor galaxies and dust-rich galaxies. The \(z=0.5\) model, on the other hand, resembles the \(z=0.0\) model but offers an improved match in the previously identified mismatched region. Regardless, the model misses a small fraction (\(<2\%\))  of dust-rich galaxies at this redshift. At \(z=1.0\), the model matches closely with the observations but similarly misses some dust-rich galaxies (\(<5\%\)). Once again, the \(z=1.5\) model closely matches the observations; however, the observations do not reflect the massive dust-poor galaxies predicted by the simulations.

We will now focus on the results from Figure \ref{sim_dsm} representing high-SFR galaxies. At \(z=0.0\), the observational data with galaxies near the minimum SFR closely aligns with the simulated galaxies, yet it misses massive high-SFR galaxies. The data shows a trend similar to that in Figure \ref{obs_dsm} at \(z=0.1\), with the simulated data failing to account for dust-rich galaxies of all sizes, which consists of \(\sim11\%\) of the total number of galaxies in this SFR range.  Additionally, while the concentration of high SFR in the observations is comparable, it is more concentrated in medium-sized galaxies than in massive ones. Once again, the \(z=0.5\) model closely resembles the \(z=0.0\) data and misses massive high-SFR galaxies. At \(z=1.0\), the model produces results consistent with those shown in Figure \ref{obs_dsm}. At \(z=1.5\), the general trend of the observational data aligns with the simulations. However, the model fails to predict the clump of small dust-rich galaxies at this redshift, consisting of \(\sim20\%\) of the total observed galaxies at this SFR range.

Figure \ref{stellar_mass_norm} coincides with a lot of the discrepancies we are seeing in these low- and high-SFR galaxies. At \(z=0.0\), we can clearly see a peak of the galaxy population in the \textsc{Simba} data around \(10^9-10^{10} M_\odot\) that aligns perfectly with the large distribution in this range in low-SFR galaxies in Figure \ref{obs_dsm}. This peak continues to \(z=0.5\) and vanishes from \(z=1.0\) onward, matching again what we see in the low-SFR figure. The massive dust-poor and rich low-SFR galaxies missing in the \textsc{Simba} data can also be seen here, as the observational data favours high-mass galaxies, while \textsc{Simba} favours low- to intermediate-mass galaxies. However, this high mass favour in the observations is likely due to the sensitivity limits in GAMA (see Section \ref{limitations}). At redshifts \(z=0.5\) to \(z=1.5\), the population of galaxies reveals a distinct difference between \textsc{Simba} and the observations, as the observations appear to favour and peak at low masses, whereas \textsc{Simba} peaks at more intermediate masses. This aligns perfectly with what we see in Figures \ref{obs_dsm} and \ref{sim_dsm}.

\section{Discussion}
\label{discussion}

Through our analysis comparing observations with simulations and using dust masses derived from \cite{Driver18} and simulated by \cite{Dave19}, we have identified several significant discrepancies in the modern simulation model \textsc{Simba}. Notably, we find a tension between the dust-rich and dust-poor galaxies modelled in \textsc{Simba} and those observed. In the following sections, we will attempt to address follow-up questions related to our conclusions, including discussing notable limitations of our datasets, comparing our work with similar analyses conducted by others, and outlining potential future research that could enhance the outcome of our results and improve our models.

\subsection{Comparison to other works}

Our understanding of how dust evolves across cosmic time has evolved rapidly over the past decade, thanks in part to the hard work done to simulate the physical processes that govern its creation, growth, and destruction. To ensure the precision of these simulations, it is crucial to compare leading models, such as \textsc{Simba}, directly with observational data and each other. This comparison helps us determine how effectively each model reproduces the observed data and where future models can improve. In this context, we compare our findings with previous works, focusing on those that have performed analyses similar to ours.

\begin{table*}[h]
    \centering
    \caption{Galaxy counts for GAMA/G10-COSMOS/3D-HST in different stellar mass, dust mass, and SFR bins.}
    \label{tab:galaxy_counts}
    \resizebox{\textwidth}{!}{%
    \begin{tabular}{cc|ccc|ccc}  
        \hline
        Redshift Bins ($z$) & $\log M_*$ [$M_{\odot}$] & \multicolumn{3}{c|}{SFR $>$ 0} & \multicolumn{3}{c}{SFR $<$ 0} \\
        \hline
        & & $\log M_d $[$M_{\odot}$] = $6-7$ & $\log M_d$[$M_{\odot}$] = $7-8$ & $\log M_d$[$M_{\odot}$] = $8-9$ & $\log M_d$[$M_{\odot}$] = $6-7$ & $\log M_d$[$M_{\odot}$] = $7-8$ & $\log M_d$[$M_{\odot}$] = $8-9$ \\
        \hline
        \multirow{3}{*}{0.0 -- 0.05} & 8.59-9.5 & 26   & 28   & 1    & 601   & 88   & 0    \\
        & 9.5-10.5 & 25   & 170   & 15   & 164   & 148   & 1   \\
        & 10.5-11.5 & 0   & 42   & 6   & 54   & 40   & 2   \\
        \hline
        \multirow{3}{*}{0.07748 -- 0.12748} & 8.59-9.5 & 513  & 492  & 57   & 1507   & 681   & 0   \\
        & 9.5-10.5 & 951  & 2753   & 392   & 2579   & 1904   & 62   \\
        & 10.5-11.5 & 45   & 833   & 272   & 1107   & 893   & 76   \\
        \hline
        \multirow{3}{*}{0.46544 -- 0.51544} 
        & 8.59-9.5 & 466  & 65   & 6   & 878   & 41   & 1   \\
        & 9.5-10.5 & 156   & 179   & 56   & 360   & 95   & 18   \\
        & 10.5-11.5 & 14   & 88   & 51   & 93   & 99   & 12   \\
        \hline
        \multirow{3}{*}{0.9677 -- 1.0177} 
        & 8.59-9.5 & 1450   & 869   & 53   & 800   & 16   & 1   \\
        & 9.5-10.5 & 413   & 886   & 178   & 496   & 218   & 24   \\
        & 10.5-11.5 & 19   & 220   & 259   & 73   & 138   & 63   \\
        \hline
        \multirow{3}{*}{1.4717 -- 1.5217} 
        & 8.59-9.5 & 655   & 471   & 260   & 568   & 15   & 0   \\
        & 9.5-10.5 & 215   & 579   & 444   & 41   & 52   & 1   \\
        & 10.5-11.5 & 11   & 116   & 242   & 1   & 13   & 7    \\
        \hline
    \end{tabular}%
    }
\end{table*}

Since our work builds directly on the findings of \cite{Li19d}, a paper that also compares with the \textsc{Simba} model, it is encouraging that our results align well with theirs. Especially at \(z=0.0\), we find that our findings agree well with the previous observations, also included in our paper, and also follow a trend similar to the simulations provided by \textsc{Simba}. We also find agreement at \(z=1.0\) with their findings and previous observations from \cite{Eales09} that \textsc{Simba} underestimates the observed DMF, especially at the higher end. With our additional data, we also found that \textsc{Simba} underestimates a large fraction of dust-poor galaxies at this redshift and similarly at \(z=0.5\). We believe that this underestimate is probably a byproduct of an inherent resolution limit \citep{Zheng21} in \textsc{Simba}, as this is consistent between \(z=0.5\) and \(z=1.5\).

The M16 (\citealp{McKinnon16}) and L25n256 (\citealp{McKinnon17a}) models are hydrodynamical dust models that track dust production, growth, and destruction up to the early universe, similar to \textsc{Simba}. We note that the L25n256 model is the successor of the M16 model that includes thermal sputtering and uses different dust growth parameters following the work by \cite{Hirashita00}. At a redshift \(z=1.0\), the high-mass end of the DMF predicted by the M16 model aligns well with our observations, though it overestimates it at \(z=0.0\). The M16 model appears to excel at assembling massive dust-rich galaxies, more so than \textsc{Simba}, which may explain its overestimation at \(z=0.0\).  On the other hand, the L25n256 model presents the opposite and matches the high-mass end at \(z=0.0\) but significantly underestimates the DMF at \(z=1.0\), similar to \textsc{Simba}.

The fiducial dust model in \cite{Popping17} is a semi-analytical model (SAM) that includes new recipes to track the production and destruction of dust up to the early universe. Despite the difference in how \textsc{Simba} and \cite{Popping17} model dust across cosmic time, both present similar dust mass functions that generally agree with each other. Due to this agreement, the fiducial model similarly underestimates the DMF at \(z=1.0\). Unlike \textsc{Simba}, the model presents a slightly worse match at \(z=0.0\) as it overestimates the DMF at the high end compared to our observations. This work also considers the dust-to-stellar mass relation as we do in Figures \ref{obs_dsm} and \ref{sim_dsm}. However, \cite{Popping17} presents their comparisons as functions, and although different, their overall shape seems to agree well with our observations at \(z=0.0\) and \(z=1.0\).

\cite{Donevski20} also compares this dust-to-stellar mass relation and uses \textsc{Simba}, similar to ours. However, they focus on earlier redshifts outside the scope of this paper, so we are unable to compare their results directly. 

Similarly to the work done by \cite{Popping17}, \cite{Triani20} uses a SAM model incorporated with their Dusty SAGE model. Their model effectively matches the dust mass functions compared to ours at \(z=0.0\) and shows good agreement with \textsc{Simba}  and may even be a better match. They also present a dust mass function at \(z=1.386\), which underestimates the function compared to our similar redshift results at \(z=1.0\) and \(z=1.5\). Additionally, their model includes a dust-to-stellar mass relation; however, like \cite{Popping17}, their comparison is based on functions, preventing a direct comparison. However, the shape of their data appears to follow a trend similar to ours.

The \cite{Hou19} model is a hydrodynamical simulation with a dust enrichment model that takes into account two different grain sizes and accounts for stellar dust production and interstellar dust processing. Unlike these other models, the proposed model disagrees with our observations at all epochs. At both \(z=0.0\) and \(z=1.0\), the model overproduces dust-rich galaxies and does not align with our observed DMFs in shape. It is possible that they produce too many massive galaxies at all redshifts in this case.

\subsection{Limitations}
\label{limitations}

Here, we discuss the limitations of our datasets to attempt to explain the discrepancies observed between \textsc{Simba} and our observations in Section \ref{Section4}.

\subsubsection{Observational limitations}
\label{observational}

As noted in Sections \ref{GAMA}-\ref{3D-HST}, our datasets vary significantly in the FIR coverage, directly affecting the dust mass estimation performed by \textsc{magphys}. The GAMA dataset provides full FIR coverage, while G10-COSMOS offers partial coverage, and the 3D-HST dataset lacks any coverage. In cases where FIR coverage is present, dust mass estimations in G10-COSMOS depend on constraints from total IR emissions (\citealp{Jin18, Magnelli24}). Meanwhile, dust mass estimations for the 3D-HST dataset depend solely on extinction measurements due to the absence of FIR data in \textsc{magphys} (\citealp{Da-Cunha08}). These indirect methods for estimating the dust mass can lead to systematic underestimations for G10-COSMOS and 3D-HST. As noted in \cite{Driver18}, the absence of FIR coverage in these surveys adds systematic error to the dust mass estimates produced by \textsc{magphys}, particularly in the earlier redshifts where G10-COSMOS and 3D-HST coverage is prevalent (see Table 6). Future research should concentrate on improving our far-infrared coverage to address this limitation.

Beyond this, these datasets are volume-limited down to specific dust mass, stellar mass, and star formation rate at low redshifts and similarly sensitivity-limited at high redshifts due to the inherent limitations of each survey (see Table 3; \citealp[]{Driver18}). These volume and sensitivity limits result in significant systematic uncertainties and limit our sample sizes (see Table 1). For example, volume limits result in a minimum dust mass that correlates with the redshift for \(z \geq 1.5\), adding restrictions to our sample (see Figure \ref{Fig2}). Furthermore, in Figure \ref{stellar_mass_norm}, we can see that GAMA and G10-COSMOS are sensitivity limited at \(z=0.1\) and \(z=1.5\), respectively,  while G10-COSMOS and 3D-HST are volume limited at \(z=0.1\) and \(z=0.5\), respectively. These limitations also introduce random errors, such as cosmic variance and Poisson errors, as well as errors from Eddington Bias (see Tables 3-6; \citealp[]{Driver18}). Improving survey volume and sensitivity is imperative to overcome this limitation, especially in earlier epochs.

Finally, while not as significant as the other limitations, AGN contamination in the data can lead to extreme estimates of stellar masses and star formation rates, with some minor impact on dust estimates.  Therefore, it was essential to remove AGN contaminants (described in Sections \ref{GAMA}, \ref{G10-COSMOS}, and \ref{3D-HST}) to avoid these effects. However, as noted in \cite{Driver18}, the removal process introduces errors (see Tables 4-5).

\subsubsection{Model caveats}
\label{caveat}
Although robust, the \textsc{Simba} model also has some caveats that affect our comparisons. We also summarise these caveats in three main points. First, the parameters that govern dust production, growth, and destruction are not well constrained, which introduces uncertainties in the model predictions (Table 1; \citealp{Li19d}). A combination of these different free parameters from what \cite{Li19d} chose may lead to a better match with the observations. Second, the model does not fully incorporate dust physics. It only models the ISM by depleting gas-phase metals and includes inactive dust particles that are only coupled to gas particles and do not change in size over time. However, the latter is more of an issue in a higher resolution simulation. Finally, the simulation of \(100h^{-1}\) Mpc lacks sufficient resolution to resolve the smallest scales of a multiphase ISM accurately. To overcome this resolution limitation, the model was adjusted such that \(\tau_{\text{ref}}\), the growth timescale of dust in the \textsc{Simba}, was modified to increase the effective gas density and the parameters governing dust destruction and condensation were fixed. \cite{Popping22}, a paper that compares multiple simulations, including \textsc{Simba}, also suggests that this time scale is too short and that modelling these effective yields requires more parameters than our simulations cannot yet resolve. Thus, constraining the free parameters in a fully resolved multiphase ISM in a higher resolution simulation will allow us to understand small-scale dust physics better and help improve the match to observational data, ushering in the need for better modelling and observations to contain those models. However, despite the presence of these dust-related caveats, we see no indication that these caveats are having a significant effect on our results.

\section{Conclusion}\label{conclusion}

We have presented dust comparisons between our observational data and the simulation model \textsc{Simba}. The \textsc{Simba} model exhibits rough agreement with our observations in DMFs and dust mass versus stellar mass across all epochs, but its limitations are apparent. From our comparisons, we concluded the following:

\begin{enumerate}

    \item[(i)] \textsc{Simba} misses dust-rich galaxies for all epochs above \(z=0.0\) (Figure \ref{Fig5}-\ref{sim_dsm}). This is consistent in low and high SFR and is likely a byproduct of the limitations of the \textsc{Simba} model at later redshifts and a byproduct of survey limitations at earlier redshifts (see Section \ref{observational}-\ref{caveat}).
    \newline
    \item[(ii)] \textsc{Simba} has a higher concentration of intermediate-mass low-SFR galaxies when compared to observations (Figure \ref{obs_dsm}-\ref{stellar_mass_norm}). This is consistent in redshifts up to \(z=0.5\). At these redshifts, \textsc{Simba} may not be able to accumulate sufficient stellar mass to form more massive dust-rich galaxies as \textsc{Simba} galaxies seem to clump around \(10^{9}-10^{10}M_\odot\). 
    \newline
    \item[(iii)] \textsc{Simba} does not accurately model low-dust mass galaxies at earlier redshifts, specifically at \(z=0.5\) and \(z=1.0\). We believe that this is a limitation in \textsc{Simba} presenting at these redshifts (Figure \ref{stellar_mass_norm}).

\end{enumerate}

Overall, with our current comparisons, we do not see specific indications that \textsc{Simba} has problems related to its implementations of dust physics, despite the limitations of the dust model itself. Rather, we predominantly see that issues arise from differences in galaxy populations, leading to the observed discrepancies in dust mass. Some of these issues could be solved with a higher resolution simulation, especially between \(z=0.5\) and \(z=1.5\), but later redshifts indicate a problem within the \textsc{Simba} model itself. As for the observational data, significant survey limitations impact some of our results, so future observations should focus on improving survey volume and FIR coverage. Several surveys promise to improve statistics here, such as WAVES for spectroscopic redshifts, Rubin Observatory's Deep Drilling Fields combined with Euclid Deep Field observations, and the ALMA-wide survey on COSMOS, which covers only the brightest galaxies and will improve statistics there.

\section*{Acknowledgements}

This research made use of Astropy, a community-developed core Python package for Astronomy \citep{Astropy-Collaboration13, Astropy-Collaboration18}.

\section*{References}

\bibliographystyle{apj}
\bibliography{references,Bibliography}

\begin{thebibliography}{80}
\expandafter\ifx\csname natexlab\endcsname\relax\def\natexlab#1{#1}\fi

\bibitem[{{Almaini} {et~al.}(2007){Almaini}, {Foucaud}, {Lane}, {Conselice}, {McLure}, {Cirasuolo}, {Dunlop}, {Smail}, \& {Simpson}}]{Almaini07}
{Almaini}, O. {et~al.} 2007, in Astronomical Society of the Pacific Conference Series, Vol. 379, Cosmic Frontiers, ed. N.~{Metcalfe} \& T.~{Shanks}, 163

\bibitem[{{Andrews} {et~al.}(2017){Andrews}, {Driver}, {Davies}, {Kafle}, {Robotham}, {Vinsen}, {Wright}, {Bland-Hawthorn}, {Bourne}, {Bremer}, {da Cunha}, {Drinkwater}, {Holwerda}, {Hopkins}, {Kelvin}, {Loveday}, {Phillipps}, \& {Wilkins}}]{Andrews17}
{Andrews}, S.~K. {et~al.} 2017, ArXiv e-prints

\bibitem[{{Andrews} {et~al.}(2017a){Andrews}, {Driver}, {Davies}, {Kafle}, {Robotham}, \& {Wright}}]{Andrews17a}
---. 2017a, MNRAS, 464 1579

\bibitem[{{Asano} {et~al.}(2013){Asano}, {Takeuchi}, {Hirashita}, \& {Nozawa}}]{Asano13}
{Asano}, R.~S. {et~al.} 2013, ArXiv e-prints

\bibitem[{{Asplund} {et~al.}(2009){Asplund}, {Grevesse}, {Sauval}, \& {Scott}}]{Asplund09}
{Asplund}, M. {et~al.} 2009, \araa, 47, 481

\bibitem[{{Astropy Collaboration} {et~al.}(2018){Astropy Collaboration}, {Price-Whelan}, {Sip{\H{o}}cz}, {G{\"u}nther}, {Lim}, {Crawford}, {Conseil}, {Shupe}, {Craig}, {Dencheva}, {Ginsburg}, {VanderPlas}, {Bradley}, {P{\'e}rez-Su{\'a}rez}, {de Val-Borro}, {Aldcroft}, {Cruz}, {Robitaille}, {Tollerud}, {Ardelean}, {Babej}, {Bach}, {Bachetti}, {Bakanov}, {Bamford}, {Barentsen}, {Barmby}, {Baumbach}, {Berry}, {Biscani}, {Boquien}, {Bostroem}, {Bouma}, {Brammer}, {Bray}, {Breytenbach}, {Buddelmeijer}, {Burke}, {Calderone}, {Cano Rodr{\'\i}guez}, {Cara}, {Cardoso}, {Cheedella}, {Copin}, {Corrales}, {Crichton}, {D'Avella}, {Deil}, {Depagne}, {Dietrich}, {Donath}, {Droettboom}, {Earl}, {Erben}, {Fabbro}, {Ferreira}, {Finethy}, {Fox}, {Garrison}, {Gibbons}, {Goldstein}, {Gommers}, {Greco}, {Greenfield}, {Groener}, {Grollier}, {Hagen}, {Hirst}, {Homeier}, {Horton}, {Hosseinzadeh}, {Hu}, {Hunkeler}, {Ivezi{\'c}}, {Jain}, {Jenness}, {Kanarek}, {Kendrew}, {Kern}, {Kerzendorf}, {Khvalko}, {King}, {Kirkby}, {Kulkarni},
  {Kumar}, {Lee}, {Lenz}, {Littlefair}, {Ma}, {Macleod}, {Mastropietro}, {McCully}, {Montagnac}, {Morris}, {Mueller}, {Mumford}, {Muna}, {Murphy}, {Nelson}, {Nguyen}, {Ninan}, {N{\"o}the}, {Ogaz}, {Oh}, {Parejko}, {Parley}, {Pascual}, {Patil}, {Patil}, {Plunkett}, {Prochaska}, {Rastogi}, {Reddy Janga}, {Sabater}, {Sakurikar}, {Seifert}, {Sherbert}, {Sherwood-Taylor}, {Shih}, {Sick}, {Silbiger}, {Singanamalla}, {Singer}, {Sladen}, {Sooley}, {Sornarajah}, {Streicher}, {Teuben}, {Thomas}, {Tremblay}, {Turner}, {Terr{\'o}n}, {van Kerkwijk}, {de la Vega}, {Watkins}, {Weaver}, {Whitmore}, {Woillez}, {Zabalza}, \& {Astropy Contributors}}]{Astropy-Collaboration18}
{Astropy Collaboration} {et~al.} 2018, \aj, 156, 123

\bibitem[{{Astropy Collaboration} {et~al.}(2013){Astropy Collaboration}, {Robitaille}, {Tollerud}, {Greenfield}, {Droettboom}, {Bray}, {Aldcroft}, {Davis}, {Ginsburg}, {Price-Whelan}, {Kerzendorf}, {Conley}, {Crighton}, {Barbary}, {Muna}, {Ferguson}, {Grollier}, {Parikh}, {Nair}, {Unther}, {Deil}, {Woillez}, {Conseil}, {Kramer}, {Turner}, {Singer}, {Fox}, {Weaver}, {Zabalza}, {Edwards}, {Azalee Bostroem}, {Burke}, {Casey}, {Crawford}, {Dencheva}, {Ely}, {Jenness}, {Labrie}, {Lim}, {Pierfederici}, {Pontzen}, {Ptak}, {Refsdal}, {Servillat}, \& {Streicher}}]{Astropy-Collaboration13}
---. 2013, \aap, 558, A33

\bibitem[{{Barger} {et~al.}(2008){Barger}, {Cowie}, \& {Wang}}]{Barger08}
{Barger}, A.~J. {et~al.} 2008, \apj, 689, 687

\bibitem[{{Beeston} {et~al.}(2018){Beeston}, {Wright}, {Maddox}, {Gomez}, {Dunne}, {Driver}, {Robotham}, {Clark}, {Vinsen}, {Takeuchi}, {Popping}, {Bourne}, {Bremer}, {Phillipps}, {Moffett}, {Baes}, {Bland-Hawthorn}, {Brough}, {De Vis}, {Eales}, {Holwerda}, {Loveday}, {Liske}, {Smith}, {Smith}, {Valiante}, {Vlahakis}, \& {Wang}}]{Beeston18}
{Beeston}, R.~A. {et~al.} 2018, \mnras, 479, 1077

\bibitem[{Bianchi \& Schneider(2007)}]{Bianchi07}
Bianchi, S., \& Schneider, R. 2007, Monthly Notices of the Royal Astronomical Society, 378, 973

\bibitem[{{Brammer} {et~al.}(2012){Brammer}, {van Dokkum}, {Franx}, {Fumagalli}, {Patel}, {Rix}, {Skelton}, {Kriek}, {Nelson}, {Schmidt}, {Bezanson}, {da Cunha}, {Erb}, {Fan}, {F{\"o}rster Schreiber}, {Illingworth}, {Labb{\'e}}, {Leja}, {Lundgren}, {Magee}, {Marchesini}, {McCarthy}, {Momcheva}, {Muzzin}, {Quadri}, {Steidel}, {Tal}, {Wake}, {Whitaker}, \& {Williams}}]{Brammer12}
{Brammer}, G.~B. {et~al.} 2012, \apjs, 200, 13

\bibitem[{{Chen} {et~al.}(2018){Chen}, {Huang}, {Liu}, {Yuan}, {Wang}, {Fan}, {Xiang}, {Zhang}, \& {Tian}}]{Chen18}
{Chen}, B.-Q. {et~al.} 2018, ArXiv e-prints

\bibitem[{{Clemens} {et~al.}(2013){Clemens}, {Pavel}, \& {Cashman}}]{Clemens13}
{Clemens}, D.~P. {et~al.} 2013, ArXiv e-prints

\bibitem[{{Cuillandre} {et~al.}(2012){Cuillandre}, {Withington}, {Hudelot}, {Goranova}, {McCracken}, {Magnard}, {Mellier}, {Regnault}, {B{\'e}toule}, {Aussel}, {Kavelaars}, {Fernique}, {Bonnarel}, {Ochsenbein}, \& {Ilbert}}]{Cuillandre12}
{Cuillandre}, J.-C.~J. {et~al.} 2012, in Society of Photo-Optical Instrumentation Engineers (SPIE) Conference Series, Vol. 8448, Observatory Operations: Strategies, Processes, and Systems IV, ed. A.~B. {Peck}, R.~L. {Seaman}, \& F.~{Comeron}, 84480M

\bibitem[{{da Cunha} {et~al.}(2008){da Cunha}, {Charlot}, \& {Elbaz}}]{Da-Cunha08}
{da Cunha}, E. {et~al.} 2008, \mnras, 388, 1595

\bibitem[{{Dav{\'e}} {et~al.}(2019){Dav{\'e}}, {Angles-Alc ´ azar}, {Narayanan}, {Qi}, {Mika}, \& {Appleby}}]{Dave19}
{Dav{\'e}}, R. {et~al.} 2019, MNRAS, 2827–2849

\bibitem[{{Davies} {et~al.}(2015){Davies}, {Robotham}, {Driver}, {Alpaslan}, {Baldry}, {Bland-Hawthorn}, {Brough}, {Brown}, {Cluver}, {Drinkwater}, {Foster}, {Grootes}, {Konstantopoulos}, {Lara-L{\'o}pez}, {L{\'o}pez-S{\'a}nchez}, {Loveday}, {Meyer}, {Moffett}, {Norberg}, {Owers}, {Popescu}, {De Propris}, {Sharp}, {Tuffs}, {Wang}, {Wilkins}, {Dunne}, {Bourne}, \& {Smith}}]{Davies15b}
{Davies}, L.~J.~M. {et~al.} 2015, \mnras, 452, 616

\bibitem[{{Davis} {et~al.}(2007){Davis}, {Mortsell}, {Sollerman}, {Becker}, {Blondin}, {Challis}, {Clocchiatti}, {Filippenko}, {Foley}, {Garnavich}, {Jha}, {Krisciunas}, {Kirshner}, {Leibundgut}, {Li}, {Matheson}, {Miknaitis}, {Pignata}, {Rest}, {Riess}, {Schmidt}, {Smith}, {Spyromilio}, {Stubbs}, {Suntzeff}, {Tonry}, \& {Wood-Vasey}}]{Davis07}
{Davis}, T.~M. {et~al.} 2007, astro-ph

\bibitem[{{Donely} {et~al.}(2012){Donely}, {Koekemoer}, {Brusa}, {Capak}, {Cardamone}, {Civano}, {Ilbert}, {Impey}, \& {Kartaltepe}}]{Donely12}
{Donely}, J.~L. {et~al.} 2012, ApJ, 748, 142

\bibitem[{{Donevski} {et~al.}(2020){Donevski}, {Lapi}, {Ma{\l}ek}, {Liu}, {G{\'o}mez-Guijarro}, {Dav{\'e}}, {Kraljic}, {Pantoni}, {Man}, {Fujimoto}, {Feltre}, {Pearson}, {Li}, \& {Narayanan}}]{Donevski20}
{Donevski}, D. {et~al.} 2020, arXiv e-prints, arXiv:2008.09995

\bibitem[{{Draine}(2003)}]{Draine03}
{Draine}, B.~T. 2003, \araa, 41, 241

\bibitem[{{Driver} {et~al.}(2018){Driver}, {Andrews}, {da Cunha}, {Davies}, {Lagos}, {Robotham}, {Vinsen}, {Wright}, {Alpaslan}, {Bland-Hawthorn}, {Bourne}, {Brough}, {Bremer}, {Cluver}, {Colless}, {Conselice}, {Dunne}, {Eales}, {Gomez}, {Holwerda}, {Hopkins}, {Kafle}, {Kelvin}, {Loveday}, {Liske}, {Maddox}, {Phillipps}, {Pimbblet}, {Rowlands}, {Sansom}, {Taylor}, {Wang}, \& {Wilkins}}]{Driver18}
{Driver}, S.~P. {et~al.} 2018, \mnras, 475, 2891

\bibitem[{{Driver} {et~al.}(2011){Driver}, {Hill}, {Kelvin}, {Robotham}, {Liske}, {Norberg}, {Baldry}, {Bamford}, {Hopkins}, {Loveday}, {Peacock}, {Andrae}, {Bland-Hawthorn}, {Brough}, {Brown}, {Cameron}, {Ching}, {Colless}, {Conselice}, {Croom}, {Cross}, {de Propris}, {Dye}, {Drinkwater}, {Ellis}, {Graham}, {Grootes}, {Gunawardhana}, {Jones}, {van Kampen}, {Maraston}, {Nichol}, {Parkinson}, {Phillipps}, {Pimbblet}, {Popescu}, {Prescott}, {Roseboom}, {Sadler}, {Sansom}, {Sharp}, {Smith}, {Taylor}, {Thomas}, {Tuffs}, {Wijesinghe}, {Dunne}, {Frenk}, {Jarvis}, {Madore}, {Meyer}, {Seibert}, {Staveley-Smith}, {Sutherland}, \& {Warren}}]{Driver11}
---. 2011, \mnras, 413, 971

\bibitem[{{Driver} {et~al.}(2009){Driver}, {Norberg}, {Baldry}, {Bamford}, {Hopkins}, {Liske}, {Loveday}, {Peacock}, {Hill}, {Kelvin}, {Robotham}, {Cross}, {Parkinson}, {Prescott}, {Conselice}, {Dunne}, {Brough}, {Jones}, {Sharp}, {van Kampen}, {Oliver}, {Roseboom}, {Bland-Hawthorn}, {Croom}, {Ellis}, {Cameron}, {Cole}, {Frenk}, {Couch}, {Alister}, {Proctor}, {De Propris}, {Doyle}, {Edmondson}, {Nichol}, {Thomas}, {Eales}, {Jarvis}, {Kuijken}, {Lahav}, {Madore}, {Seibert}, {Meyer}, {Staveley-Smith}, {Phillipps}, {Popescu}, {Sansom}, {Sutherland}, {Tuffs}, \& {Warren}}]{Driver09}
---. 2009, Astronomy and Geophysics, 50, 050000

\bibitem[{{Dunne} {et~al.}(2011){Dunne}, {Gomez}, {da Cunha}, {Charlot}, {Dye}, {Eales}, {Maddox}, {Rowlands}, {Smith}, {Auld}, {Baes}, {Bonfield}, {Bourne}, {Buttiglione}, {Cava}, {Clements}, {Coppin}, {Cooray}, {Dariush}, {de Zotti}, {Driver}, {Fritz}, {Geach}, {Hopwood}, {Ibar}, {Ivison}, {Jarvis}, {Kelvin}, {Pascale}, {Pohlen}, {Popescu}, {Rigby}, {Robotham}, {Rodighiero}, {Sansom}, {Serjeant}, {Temi}, {Thompson}, {Tuffs}, {van der Werf}, \& {Vlahakis}}]{Dunne11}
{Dunne}, L. {et~al.} 2011, \mnras, 1395

\bibitem[{{Dwek}(1998)}]{Dwek98}
{Dwek}, E. 1998, \apj, 501, 643

\bibitem[{{Dwek} \& {Scalo}(1980)}]{Dwek80}
{Dwek}, E., \& {Scalo}, J.~M. 1980, \apj, 239, 193

\bibitem[{{Eales} {et~al.}(2009){Eales}, {Chapin}, {Devlin}, {Dye}, {Halpern}, {Hughes}, {Marsden}, {Mauskopf}, {Moncelsi}, {Netterfield}, {Pascale}, {Patanchon}, {Raymond}, {Rex}, {Scott}, {Semisch}, {Siana}, {Truch}, \& {Viero}}]{Eales09}
{Eales}, S. {et~al.} 2009, \apj, 707, 1779

\bibitem[{{Eales} {et~al.}(2010){Eales}, {Cooray}, \& {Dunne}}]{Eales10c}
---. 2010, PASP, 122, 499

\bibitem[{{Eales} \& {Ward}(2024)}]{Eales24}
{Eales}, S., \& {Ward}, B. 2024, \mnras, 529, 1130

\bibitem[{{Emerson} \& {Sutherland}(2002)}]{Emerson02}
{Emerson}, J.~P., \& {Sutherland}, W. 2002, in Society of Photo-Optical Instrumentation Engineers (SPIE) Conference Series, Vol. 4836, Survey and Other Telescope Technologies and Discoveries, ed. J.~A. {Tyson} \& S.~{Wolff}, 35--42

\bibitem[{{Ferrarotti} \& {Gail}(2006)}]{Ferrarotti06}
{Ferrarotti}, A.~S., \& {Gail}, H.~P. 2006, A\&A, 447, 553

\bibitem[{{Griffin} {et~al.}(2010){Griffin}, {Abergel}, {Abreu}, {Ade}, {Andr{\'e}}, {Augueres}, {Babbedge}, {Bae}, {Baillie}, {Baluteau}, {Barlow}, {Bendo}, {Benielli}, {Bock}, {Bonhomme}, {Brisbin}, {Brockley-Blatt}, {Caldwell}, {Cara}, {Castro-Rodriguez}, {Cerulli}, {Chanial}, {Chen}, {Clark}, {Clements}, {Clerc}, {Coker}, {Communal}, {Conversi}, {Cox}, {Crumb}, {Cunningham}, {Daly}, {Davis}, {de Antoni}, {Delderfield}, {Devin}, {di Giorgio}, {Didschuns}, {Dohlen}, {Donati}, {Dowell}, {Dowell}, {Duband}, {Dumaye}, {Emery}, {Ferlet}, {Ferrand}, {Fontignie}, {Fox}, {Franceschini}, {Frerking}, {Fulton}, {Garcia}, {Gastaud}, {Gear}, {Glenn}, {Goizel}, {Griffin}, {Grundy}, {Guest}, {Guillemet}, {Hargrave}, {Harwit}, {Hastings}, {Hatziminaoglou}, {Herman}, {Hinde}, {Hristov}, {Huang}, {Imhof}, {Isaak}, {Israelsson}, {Ivison}, {Jennings}, {Kiernan}, {King}, {Lange}, {Latter}, {Laurent}, {Laurent}, {Leeks}, {Lellouch}, {Levenson}, {Li}, {Li}, {Lilienthal}, {Lim}, {Liu}, {Lu}, {Madden}, {Mainetti}, {Marliani},
  {McKay}, {Mercier}, {Molinari}, {Morris}, {Moseley}, {Mulder}, {Mur}, {Naylor}, {Nguyen}, {O'Halloran}, {Oliver}, {Olofsson}, {Olofsson}, {Orfei}, {Page}, {Pain}, {Panuzzo}, {Papageorgiou}, {Parks}, {Parr-Burman}, {Pearce}, {Pearson}, {P{\'e}rez-Fournon}, {Pinsard}, {Pisano}, {Podosek}, {Pohlen}, {Polehampton}, {Pouliquen}, {Rigopoulou}, {Rizzo}, {Roseboom}, {Roussel}, {Rowan-Robinson}, {Rownd}, {Saraceno}, {Sauvage}, {Savage}, {Savini}, {Sawyer}, {Scharmberg}, {Schmitt}, {Schneider}, {Schulz}, {Schwartz}, {Shafer}, {Shupe}, {Sibthorpe}, {Sidher}, {Smith}, {Smith}, {Smith}, {Spencer}, {Stobie}, {Sudiwala}, {Sukhatme}, {Surace}, {Stevens}, {Swinyard}, {Trichas}, {Tourette}, {Triou}, {Tseng}, {Tucker}, {Turner}, {Vaccari}, {Valtchanov}, {Vigroux}, {Virique}, {Voellmer}, {Walker}, {Ward}, {Waskett}, {Weilert}, {Wesson}, {White}, {Whitehouse}, {Wilson}, {Winter}, {Woodcraft}, {Wright}, {Xu}, {Zavagno}, {Zemcov}, {Zhang}, \& {Zonca}}]{Griffin10}
{Griffin}, M.~J. {et~al.} 2010, \aap, 518, L3+

\bibitem[{{Grogin} {et~al.}(2011){Grogin}, {Kocevski}, {Faber}, {Ferguson}, {Koekemoer}, {Riess}, {Acquaviva}, {Alexander}, {Almaini}, {Ashby}, {Barden}, {Bell}, {Bournaud}, {Brown}, {Caputi}, {Casertano}, {Cassata}, {Castellano}, {Challis}, {Chary}, {Cheung}, {Cirasuolo}, {Conselice}, {Roshan Cooray}, {Croton}, {Daddi}, {Dahlen}, {Dav{\'e}}, {de Mello}, {Dekel}, {Dickinson}, {Dolch}, {Donley}, {Dunlop}, {Dutton}, {Elbaz}, {Fazio}, {Filippenko}, {Finkelstein}, {Fontana}, {Gardner}, {Garnavich}, {Gawiser}, {Giavalisco}, {Grazian}, {Guo}, {Hathi}, {H{\"a}ussler}, {Hopkins}, {Huang}, {Huang}, {Jha}, {Kartaltepe}, {Kirshner}, {Koo}, {Lai}, {Lee}, {Li}, {Lotz}, {Lucas}, {Madau}, {McCarthy}, {McGrath}, {McIntosh}, {McLure}, {Mobasher}, {Moustakas}, {Mozena}, {Nandra}, {Newman}, {Niemi}, {Noeske}, {Papovich}, {Pentericci}, {Pope}, {Primack}, {Rajan}, {Ravindranath}, {Reddy}, {Renzini}, {Rix}, {Robaina}, {Rodney}, {Rosario}, {Rosati}, {Salimbeni}, {Scarlata}, {Siana}, {Simard}, {Smidt}, {Somerville}, {Spinrad},
  {Straughn}, {Strolger}, {Telford}, {Teplitz}, {Trump}, {van der Wel}, {Villforth}, {Wechsler}, {Weiner}, {Wiklind}, {Wild}, {Wilson}, {Wuyts}, {Yan}, \& {Yun}}]{Grogin11}
{Grogin}, N.~A. {et~al.} 2011, \apjs, 197, 35

\bibitem[{{Hensley} \& {Draine}(2023)}]{Hensley23}
{Hensley}, B.~S., \& {Draine}, B.~T. 2023, \apj, 948, 55

\bibitem[{{Hirashita}(2000)}]{Hirashita00}
{Hirashita}, H. 2000, PASJ, 52, 585

\bibitem[{{Hopkins} \& {Lee}(2015)}]{Hopkins15}
{Hopkins}, P.~F., \& {Lee}, H. 2015, ArXiv e-prints

\bibitem[{{Hou} {et~al.}(2019){Hou}, {Aoyama}, {Hirashita}, {Nagamine}, \& {Shimizu}}]{Hou19}
{Hou}, K.-C. {et~al.} 2019, MNRAS, 485, 1727

\bibitem[{{Hurley} {et~al.}(2017){Hurley}, {Oliver}, {Betacourt}, {Clarke}, {Cowley}, {Duivenvoorden}, {Farrah}, {Griffin}, \& {Lacey}}]{Hurley17}
{Hurley}, P.~D. {et~al.} 2017, MNRAS, 464, 885

\bibitem[{{Jarvis} {et~al.}(2013){Jarvis}, {Bonfield}, {Bruce}, {Geach}, {McAlpine}, {McLure}, {Gonz{\'a}lez-Solares}, {Irwin}, {Lewis}, {Yoldas}, {Andreon}, {Cross}, {Emerson}, {Dalton}, {Dunlop}, {Hodgkin}, {Le}, {Karouzos}, {Meisenheimer}, {Oliver}, {Rawlings}, {Simpson}, {Smail}, {Smith}, {Sullivan}, {Sutherland}, {White}, \& {Zwart}}]{Jarvis13}
{Jarvis}, M.~J. {et~al.} 2013, \mnras, 428, 1281

\bibitem[{{Jin} {et~al.}(2018){Jin}, {Daddi}, {Liu}, {Smolcic}, {Schinnerer}, {Calabr{\`o}}, {Gu}, {Delhaize}, {Delvecchio}, {Gao}, {Salvato}, {Puglisi}, {Dickinson}, {Bertoldi}, {Sargent}, {Novak}, {Magdis}, {Aretxaga}, {Wilson}, \& {Capak}}]{Jin18}
{Jin}, S. {et~al.} 2018, ArXiv e-prints

\bibitem[{{Laigle} {et~al.}(2016){Laigle}, {McCracken}, {Ilbert}, {Hsieh}, {Davidzon}, {Capak}, {Hasinger}, {Silverman}, {Pichon}, {Coupon}, {Aussel}, {Le Borgne}, {Caputi}, {Cassata}, {Chang}, {Civano}, {Dunlop}, {Fynbo}, {kartaltepe}, {Koekemoer}, {Le Fevre}, {Le Floc'h}, {Leauthaud}, {Lilly}, {Lin}, {Marchesi}, {Milvang-Jensen}, {Salvato}, {Sanders}, {Scoville}, {Smolcic}, {Stockmann}, {Taniguchi}, {Tasca}, {Toft}, {Vaccari}, \& {Zabl}}]{Laigle16}
{Laigle}, C. {et~al.} 2016, ArXiv e-prints

\bibitem[{{Leśniewska, Aleksandra} \& {Michałowski, Michał Jerzy}(2019)}]{Leśniewska19}
{Leśniewska, Aleksandra}, \& {Michałowski, Michał Jerzy}. 2019, A\&A, 624, L13

\bibitem[{{Li} {et~al.}(2019){Li}, {Narayanan}, \& {Dav{\'e}}}]{Li19d}
{Li}, Q. {et~al.} 2019, arXiv e-prints

\bibitem[{{Liske} {et~al.}(2015){Liske}, {Baldry}, {Driver}, {Tuffs}, {Alpaslan}, {Andrae}, {Brough}, {Cluver}, {Grootes}, {Gunawardhana}, {Kelvin}, {Loveday}, {Robotham}, {Taylor}, {Bamford}, {Bland-Hawthorn}, {Brown}, {Drinkwater}, {Hopkins}, {Meyer}, {Norberg}, {Peacock}, {Agius}, {Andrews}, {Bauer}, {Ching}, {Colless}, {Conselice}, {Croom}, {Davies}, {De Propris}, {Dunne}, {Eardley}, {Ellis}, {Foster}, {Frenk}, {H{\"a}u{\ss}ler}, {Holwerda}, {Howlett}, {Ibarra}, {Jarvis}, {Jones}, {Kafle}, {Lacey}, {Lange}, {Lara-L{\'o}pez}, {L{\'o}pez-S{\'a}nchez}, {Maddox}, {Madore}, {McNaught-Roberts}, {Moffett}, {Nichol}, {Owers}, {Palamara}, {Penny}, {Phillipps}, {Pimbblet}, {Popescu}, {Prescott}, {Proctor}, {Sadler}, {Sansom}, {Seibert}, {Sharp}, {Sutherland}, {V{\'a}zquez-Mata}, {van Kampen}, {Wilkins}, {Williams}, \& {Wright}}]{Liske15}
{Liske}, J. {et~al.} 2015, \mnras, 452, 2087

\bibitem[{{Magnelli} {et~al.}(2024){Magnelli}, {Adscheid}, {Wang}, {Ciesla}, {Daddi}, {Delvecchio}, {Elbaz}, {Fudamoto}, {Fukushima}, {Franco}, {G{\'o}mez-Guijarro}, {Gruppioni}, {Jim{\'e}nez-Andrade}, {Liu}, {Oesch}, {Schinnerer}, \& {Traina}}]{Magnelli24}
{Magnelli}, B. {et~al.} 2024, arXiv e-prints, arXiv:2405.18086

\bibitem[{{Martin} {et~al.}(2005){Martin}, {Fanson}, {Schiminovich}, {Morrissey}, {Friedman}, {Barlow}, {Conrow}, {Grange}, {Jelinsky}, {Milliard}, {Siegmund}, {Bianchi}, {Byun}, {Donas}, {Forster}, {Heckman}, {Lee}, {Madore}, {Malina}, {Neff}, {Rich}, {Small}, {Surber}, {Szalay}, {Welsh}, \& {Wyder}}]{Martin05}
{Martin}, D.~C. {et~al.} 2005, \apjl, 619, L1

\bibitem[{{McKee}(1989)}]{McKee89}
{McKee}, C. 1989, Interstellar Dust: Proceedings of the 135th Symposium of the International Astronomical Union,, 431

\bibitem[{{McKee} {et~al.}(1987){McKee}, {Hollenbach}, {Seab}, \& {Tielens}}]{McKee87}
{McKee}, C.~F. {et~al.} 1987, \apj, 318, 674

\bibitem[{{McKinnon} {et~al.}(2017){McKinnon}, {Torrey}, {Vogelsberger}, {Hayward}, \& {Marinacci}}]{McKinnon17a}
{McKinnon}, R. {et~al.} 2017, MNRAS, 468, 1505

\bibitem[{{McKinnon} {et~al.}(2016){McKinnon}, {Torrey}, {Vogelsberger}, {Hayward}, \& {Marinacci}}]{McKinnon16}
---. 2016, ArXiv e-prints

\bibitem[{{Momcheva} {et~al.}(2016){Momcheva}, {Brammer}, {van Dokkum}, {Skelton}, {Whitaker}, {Nelson}, {Fumagalli}, {Maseda}, {Leja}, {Franx}, {Rix}, {Bezanson}, {Da Cunha}, {Dickey}, {F{\"o}rster Schreiber}, {Illingworth}, {Kriek}, {Labb{\'e}}, {Ulf Lange}, {Lundgren}, {Magee}, {Marchesini}, {Oesch}, {Pacifici}, {Patel}, {Price}, {Tal}, {Wake}, {van der Wel}, \& {Wuyts}}]{Momcheva16a}
{Momcheva}, I.~G. {et~al.} 2016, \apjs, 225, 27

\bibitem[{{Nonino} {et~al.}(2009){Nonino}, {Dickinson}, {Rosati}, {Grazian}, {Reddy}, {Cristiani}, {Giavalisco}, {Kuntschner}, {Vanzella}, {Daddi}, {Fosbury}, \& {Cesarsky}}]{Nonino09}
{Nonino}, M. {et~al.} 2009, \apjs, 183, 244

\bibitem[{{Oliver} {et~al.}(2012){Oliver}, {Bock}, {Altieri}, {Amblard}, {Arumugam}, {Aussel}, {Babbedge}, {Beelen}, {B{\'e}thermin}, {Blain}, {Boselli}, {Bridge}, {Brisbin}, {Buat}, {Burgarella}, {Castro-Rodr{\'\i}guez}, {Cava}, {Chanial}, {Cirasuolo}, {Clements}, {Conley}, {Conversi}, {Cooray}, {Dowell}, {Dubois}, {Dwek}, {Dye}, {Eales}, {Elbaz}, {Farrah}, {Feltre}, {Ferrero}, {Fiolet}, {Fox}, {Franceschini}, {Gear}, {Giovannoli}, {Glenn}, {Gong}, {Gonz{\'a}lez Solares}, {Griffin}, {Halpern}, {Harwit}, {Hatziminaoglou}, {Heinis}, {Hurley}, {Hwang}, {Hyde}, {Ibar}, {Ilbert}, {Isaak}, {Ivison}, {Lagache}, {Le Floc'h}, {Levenson}, {Faro}, {Lu}, {Madden}, {Maffei}, {Magdis}, {Mainetti}, {Marchetti}, {Marsden}, {Marshall}, {Mortier}, {Nguyen}, {O'Halloran}, {Omont}, {Page}, {Panuzzo}, {Papageorgiou}, {Patel}, {Pearson}, {P{\'e}rez-Fournon}, {Pohlen}, {Rawlings}, {Raymond}, {Rigopoulou}, {Riguccini}, {Rizzo}, {Rodighiero}, {Roseboom}, {Rowan-Robinson}, {S{\'a}nchez Portal}, {Schulz}, {Scott}, {Seymour}, {Shupe},
  {Smith}, {Stevens}, {Symeonidis}, {Trichas}, {Tugwell}, {Vaccari}, {Valtchanov}, {Vieira}, {Viero}, {Vigroux}, {Wang}, {Ward}, {Wardlow}, {Wright}, {Xu}, \& {Zemcov}}]{Oliver12}
{Oliver}, S.~J. {et~al.} 2012, \mnras, 424, 1614

\bibitem[{Omukai {et~al.}(2005)Omukai, Tsuribe, Schneider, \& Ferrara}]{Omukai05}
Omukai, K. {et~al.} 2005, The Astrophysical Journal, 626, 627

\bibitem[{{Parente} {et~al.}(2022){Parente}, {Ragone-Figueroa}, {Granato}, {Borgani}, {Murante}, {Valentini}, {Bressan}, \& {Lapi}}]{Parente22a}
{Parente}, M. {et~al.} 2022, \mnras, 515, 2053

\bibitem[{{Parente} {et~al.}(2023){Parente}, {Ragone-Figueroa}, {Granato}, \& {Lapi}}]{Parente23}
---. 2023, \mnras, 521, 6105

\bibitem[{{Pilbratt}(2003)}]{Pilbratt03}
{Pilbratt}, G.~L. 2003, in Society of Photo-Optical Instrumentation Engineers (SPIE) Conference Series, Vol. 4850, IR Space Telescopes and Instruments, ed. J.~C. {Mather}, 586--597

\bibitem[{{Planck Collaboration} {et~al.}(2016){Planck Collaboration}, {Ade}, {Aghanim}, {Arnaud}, {Ashdown}, {Aumont}, {Baccigalupi}, {Banday}, {Barreiro}, {Bartlett}, {Bartolo}, {Battaner}, {Battye}, {Benabed}, {Beno{\^\i}t}, {Benoit-L{\'e}vy}, {Bernard}, {Bersanelli}, {Bielewicz}, {Bock}, {Bonaldi}, {Bonavera}, {Bond}, {Borrill}, {Bouchet}, {Boulanger}, {Bucher}, {Burigana}, {Butler}, {Calabrese}, {Cardoso}, {Catalano}, {Challinor}, {Chamballu}, {Chary}, {Chiang}, {Chluba}, {Christensen}, {Church}, {Clements}, {Colombi}, {Colombo}, {Combet}, {Coulais}, {Crill}, {Curto}, {Cuttaia}, {Danese}, {Davies}, {Davis}, {de Bernardis}, {de Rosa}, {de Zotti}, {Delabrouille}, {D{\'e}sert}, {Di Valentino}, {Dickinson}, {Diego}, {Dolag}, {Dole}, {Donzelli}, {Dor{\'e}}, {Douspis}, {Ducout}, {Dunkley}, {Dupac}, {Efstathiou}, {Elsner}, {En{\ss}lin}, {Eriksen}, {Farhang}, {Fergusson}, {Finelli}, {Forni}, {Frailis}, {Fraisse}, {Franceschi}, {Frejsel}, {Galeotta}, {Galli}, {Ganga}, {Gauthier}, {Gerbino}, {Ghosh}, {Giard},
  {Giraud-H{\'e}raud}, {Giusarma}, {Gjerl{\o}w}, {Gonz{\'a}lez-Nuevo}, {G{\'o}rski}, {Gratton}, {Gregorio}, {Gruppuso}, {Gudmundsson}, {Hamann}, {Hansen}, {Hanson}, {Harrison}, {Helou}, {Henrot-Versill{\'e}}, {Hern{\'a}ndez-Monteagudo}, {Herranz}, {Hildebrand t}, {Hivon}, {Hobson}, {Holmes}, {Hornstrup}, {Hovest}, {Huang}, {Huffenberger}, {Hurier}, {Jaffe}, {Jaffe}, {Jones}, {Juvela}, {Keih{\"a}nen}, {Keskitalo}, {Kisner}, {Kneissl}, {Knoche}, {Knox}, {Kunz}, {Kurki-Suonio}, {Lagache}, {L{\"a}hteenm{\"a}ki}, {Lamarre}, {Lasenby}, {Lattanzi}, {Lawrence}, {Leahy}, {Leonardi}, {Lesgourgues}, {Levrier}, {Lewis}, {Liguori}, {Lilje}, {Linden-V{\o}rnle}, {L{\'o}pez-Caniego}, {Lubin}, {Mac{\'\i}as-P{\'e}rez}, {Maggio}, {Maino}, {Mandolesi}, {Mangilli}, {Marchini}, {Maris}, {Martin}, {Martinelli}, {Mart{\'\i}nez-Gonz{\'a}lez}, {Masi}, {Matarrese}, {McGehee}, {Meinhold}, {Melchiorri}, {Melin}, {Mendes}, {Mennella}, {Migliaccio}, {Millea}, {Mitra}, {Miville-Desch{\^e}nes}, {Moneti}, {Montier}, {Morgante}, {Mortlock},
  {Moss}, {Munshi}, {Murphy}, {Naselsky}, {Nati}, {Natoli}, {Netterfield}, {N{\o}rgaard-Nielsen}, {Noviello}, {Novikov}, {Novikov}, {Oxborrow}, {Paci}, {Pagano}, {Pajot}, {Paladini}, {Paoletti}, {Partridge}, {Pasian}, {Patanchon}, {Pearson}, {Perdereau}, {Perotto}, {Perrotta}, {Pettorino}, {Piacentini}, {Piat}, {Pierpaoli}, {Pietrobon}, {Plaszczynski}, {Pointecouteau}, {Polenta}, {Popa}, {Pratt}, {Pr{\'e}zeau}, {Prunet}, {Puget}, {Rachen}, {Reach}, {Rebolo}, {Reinecke}, {Remazeilles}, {Renault}, {Renzi}, {Ristorcelli}, {Rocha}, {Rosset}, {Rossetti}, {Roudier}, {Rouill{\'e} d'Orfeuil}, {Rowan-Robinson}, {Rubi{\~n}o-Mart{\'\i}n}, {Rusholme}, {Said}, {Salvatelli}, {Salvati}, {Sandri}, {Santos}, {Savelainen}, {Savini}, {Scott}, {Seiffert}, {Serra}, {Shellard}, {Spencer}, {Spinelli}, {Stolyarov}, {Stompor}, {Sudiwala}, {Sunyaev}, {Sutton}, {Suur-Uski}, {Sygnet}, {Tauber}, {Terenzi}, {Toffolatti}, {Tomasi}, {Tristram}, {Trombetti}, {Tucci}, {Tuovinen}, {T{\"u}rler}, {Umana}, {Valenziano}, {Valiviita}, {Van Tent},
  {Vielva}, {Villa}, {Wade}, {Wandelt}, {Wehus}, {White}, {White}, {Wilkinson}, {Yvon}, {Zacchei}, \& {Zonca}}]{Planck-Collaboration16}
{Planck Collaboration} {et~al.} 2016, \aap, 594, A13

\bibitem[{{Poglitsch} {et~al.}(2010){Poglitsch}, {Waelkens}, {Geis}, {Feuchtgruber}, {Vandenbussche}, {Rodriguez}, {Krause}, {Renotte}, {van Hoof}, {Saraceno}, {Cepa}, {Kerschbaum}, {Agn{\`e}se}, {Ali}, {Altieri}, {Andreani}, {Augueres}, {Balog}, {Barl}, {Bauer}, {Belbachir}, {Benedettini}, {Billot}, {Boulade}, {Bischof}, {Blommaert}, {Callut}, {Cara}, {Cerulli}, {Cesarsky}, {Contursi}, {Creten}, {De Meester}, {Doublier}, {Doumayrou}, {Duband}, {Exter}, {Genzel}, {Gillis}, {Gr{\"o}zinger}, {Henning}, {Herreros}, {Huygen}, {Inguscio}, {Jakob}, {Jamar}, {Jean}, {de Jong}, {Katterloher}, {Kiss}, {Klaas}, {Lemke}, {Lutz}, {Madden}, {Marquet}, {Martignac}, {Mazy}, {Merken}, {Montfort}, {Morbidelli}, {M{\"u}ller}, {Nielbock}, {Okumura}, {Orfei}, {Ottensamer}, {Pezzuto}, {Popesso}, {Putzeys}, {Regibo}, {Reveret}, {Royer}, {Sauvage}, {Schreiber}, {Stegmaier}, {Schmitt}, {Schubert}, {Sturm}, {Thiel}, {Tofani}, {Vavrek}, {Wetzstein}, {Wieprecht}, \& {Wiezorrek}}]{Poglitsch10}
{Poglitsch}, A. {et~al.} 2010, \aap, 518, L2+

\bibitem[{{Popping} {et~al.}(2017){Popping}, {Puglisi}, \& {Norman}}]{Popping17}
{Popping}, G. {et~al.} 2017, ArXiv e-prints

\bibitem[{{Popping} \& {Péroux}(2022)}]{Popping22}
{Popping}, G., \& {Péroux}, C. 2022, MNRAS, 513, 1531

\bibitem[{{Ragone-Figueroa} {et~al.}(2024){Ragone-Figueroa}, {Granato}, {Parente}, {Murante}, {Valentini}, {Borgani}, \& {Maio}}]{Ragone-Figueroa24}
{Ragone-Figueroa}, C. {et~al.} 2024, arXiv e-prints, arXiv:2407.06269

\bibitem[{{Sanders} {et~al.}(2007){Sanders}, {Salvato}, {Aussel}, {Ilbert}, {Scoville}, {Surace}, {Frayer}, {Sheth}, {Helou}, {Brooke}, {Bhattacharya}, {Yan}, {Kartaltepe}, {Barnes}, {Blain}, {Calzetti}, {Capak}, {Carilli}, {Carollo}, {Comastri}, {Daddi}, {Ellis}, {Elvis}, {Fall}, {Franceschini}, {Giavalisco}, {Hasinger}, {Impey}, {Koekemoer}, {Le F{\`e}vre}, {Lilly}, {Liu}, {McCracken}, {Mobasher}, {Renzini}, {Rich}, {Schinnerer}, {Shopbell}, {Taniguchi}, {Thompson}, {Urry}, \& {Williams}}]{Sanders07}
{Sanders}, D.~B. {et~al.} 2007, \apjs, 172, 86

\bibitem[{{Santini} {et~al.}(2014){Santini}, {Maiolino}, \& {Magnelli}}]{Santini14}
{Santini}, P. {et~al.} 2014, A\&A

\bibitem[{Schneider {et~al.}(2006)Schneider, Omukai, Inoue, \& Ferrara}]{Schneider06}
Schneider, R. {et~al.} 2006, Monthly Notices of the Royal Astronomical Society, 369, 1437

\bibitem[{{Scoville} {et~al.}(2007{\natexlab{a}}){Scoville}, {Abraham}, {Aussel}, {Barnes}, {Benson}, {Blain}, {Calzetti}, {Comastri}, {Capak}, {Carilli}, {Carlstrom}, {Carollo}, {Colbert}, {Daddi}, {Ellis}, {Elvis}, {Ewald}, {Fall}, {Franceschini}, {Giavalisco}, {Green}, {Griffiths}, {Guzzo}, {Hasinger}, {Impey}, {Kneib}, {Koda}, {Koekemoer}, {Lefevre}, {Lilly}, {Liu}, {McCracken}, {Massey}, {Mellier}, {Miyazaki}, {Mobasher}, {Mould}, {Norman}, {Refregier}, {Renzini}, {Rhodes}, {Rich}, {Sanders}, {Schiminovich}, {Schinnerer}, {Scodeggio}, {Sheth}, {Shopbell}, {Taniguchi}, {Tyson}, {Urry}, {Van Waerbeke}, {Vettolani}, {White}, \& {Yan}}]{Scoville07b}
{Scoville}, N. {et~al.} 2007{\natexlab{a}}, \apjs, 172, 38

\bibitem[{{Scoville} {et~al.}(2007{\natexlab{b}}){Scoville}, {Aussel}, {Brusa}, {Capak}, {Carollo}, {Elvis}, {Giavalisco}, {Guzzo}, {Hasinger}, {Impey}, {Kneib}, {LeFevre}, {Lilly}, {Mobasher}, {Renzini}, {Rich}, {Sanders}, {Schinnerer}, {Schminovich}, {Shopbell}, {Taniguchi}, \& {Tyson}}]{Scoville07a}
---. 2007{\natexlab{b}}, \apjs, 172, 1

\bibitem[{{Seab} \& {Shull}(1983)}]{Seab83}
{Seab}, C.~G., \& {Shull}, J.~M. 1983, \apj, 275, 652

\bibitem[{{Seymour} {et~al.}(2008){Seymour}, {Dwelly}, {Moss}, {McHardy}, {Zoghbi}, {Rieke}, {Page}, {Hopkins}, \& {Loaring}}]{Seymour08}
{Seymour}, N. {et~al.} 2008, MNRAS, 386, 1695

\bibitem[{{Shirley} {et~al.}(2021){Shirley}, {Duncan}, {Campos Varillas}, {Hurley}, {Ma{\l}ek}, {Roehlly}, {Smith}, {Aussel}, {Bakx}, {Buat}, {Burgarella}, {Christopher}, {Duivenvoorden}, {Eales}, {Efstathiou}, {Gonz{\'a}lez Solares}, {Griffin}, {Jarvis}, {Faro}, {Marchetti}, {McCheyne}, {Papadopoulos}, {Penner}, {Pons}, {Prescott}, {Rigby}, {Rottgering}, {Saxena}, {Scudder}, {Vaccari}, {Wang}, \& {Oliver}}]{Shirley21}
{Shirley}, R. {et~al.} 2021, \mnras, 507, 129

\bibitem[{{Taniguchi} {et~al.}(2007){Taniguchi}, {Scoville}, {Murayama}, {Sanders}, {Mobasher}, {Aussel}, {Capak}, {Ajiki}, {Miyazaki}, {Komiyama}, {Shioya}, {Nagao}, {Sasaki}, {Koda}, {Carilli}, {Giavalisco}, {Guzzo}, {Hasinger}, {Impey}, {LeFevre}, {Lilly}, {Renzini}, {Rich}, {Schinnerer}, {Shopbell}, {Kaifu}, {Karoji}, {Arimoto}, {Okamura}, \& {Ohta}}]{Taniguchi07}
{Taniguchi}, Y. {et~al.} 2007, \apjs, 172, 9

\bibitem[{Todini \& Ferrara(2001)}]{Todini01}
Todini, P., \& Ferrara, A. 2001, Monthly Notices of the Royal Astronomical Society, 325, 726

\bibitem[{{Triani} {et~al.}(2020){Triani}, {Sinha}, {Croton}, {Pacifici}, \& {Dwek}}]{Triani20}
{Triani}, D.~P. {et~al.} 2020, \mnras

\bibitem[{{Vijayan} {et~al.}(2019){Vijayan}, {Clay}, {Thomas}, {Yates}, {Wilkins}, \& {Henriques}}]{Vijayan19}
{Vijayan}, A.~P. {et~al.} 2019, arXiv e-prints

\bibitem[{Wakelam {et~al.}(2017)Wakelam, Bron, Cazaux, Dulieu, Gry, Guillard, Habart, Hornekær, Morisset, Nyman, Pirronello, Price, Valdivia, Vidali, \& Watanabe}]{Wakelam17}
Wakelam, V. {et~al.} 2017, Molecular Astrophysics, 9, 1

\bibitem[{{Werner} {et~al.}(2004){Werner}, {Roellig}, {Low}, {Rieke}, {Rieke}, {Hoffmann}, {Young}, {Houck}, {Brandl}, {Fazio}, {Hora}, {Gehrz}, {Helou}, {Soifer}, {Stauffer}, {Keene}, {Eisenhardt}, {Gallagher}, {Gautier}, {Irace}, {Lawrence}, {Simmons}, {Van Cleve}, {Jura}, {Wright}, \& {Cruikshank}}]{Werner04}
{Werner}, M.~W. {et~al.} 2004, \apjs, 154, 1

\bibitem[{{Wright} {et~al.}(2016){Wright}, {Robotham}, {Bourne}, {Driver}, {Dunne}, {Maddox}, {Alpaslan}, {Andrews}, {Bauer}, {Bland-Hawthorn}, {Brough}, {Brown}, {Clarke}, {Cluver}, {Davies}, {Grootes}, {Holwerda}, {Hopkins}, {Jarrett}, {Kafle}, {Lange}, {Liske}, {Loveday}, {Moffett}, {Norberg}, {Popescu}, {Smith}, {Taylor}, {Tuffs}, {Wang}, \& {Wilkins}}]{Wright16}
{Wright}, A.~H. {et~al.} 2016, \mnras, 460, 765

\bibitem[{{Yates} {et~al.}(2024){Yates}, {Hendriks}, {Vijayan}, {Izzard}, {Thomas}, \& {Das}}]{Yates24}
{Yates}, R.~M. {et~al.} 2024, \mnras, 527, 6292

\bibitem[{{Zheng} {et~al.}(2021){Zheng}, {Dave}, {Wild}, \& {Rodr{\'\i}guez Montero}}]{Zheng21}
{Zheng}, Y. {et~al.} 2021, arXiv e-prints, arXiv:2110.01935

\end{thebibliography}

\end{document}